\documentclass[a4paper,11pt]{article}
\usepackage{jheppub}
\usepackage{dcolumn}
\usepackage{bm}
\usepackage{graphicx}
\usepackage{epstopdf}
\usepackage{subfigure}
\usepackage{tikz}
\usepackage{siunitx}
\usepackage{float}
\usepackage{booktabs}
\usepackage{url}
\usepackage[utf8]{inputenc}
\usepackage{color}
\usepackage{makecell}
\usepackage{threeparttable}
\usepackage{tikz}
\usepackage[compat=1.1.0]{tikz-feynman}
\usepackage{tikz-feynman}
\usepackage{geometry}
\usepackage{slashed}
\newcommand\blfootnote[1]{%
  \begingroup
  \renewcommand\thefootnote{}\footnote{#1}%
  \addtocounter{footnote}{-1}%
  \endgroup
}

\usetikzlibrary{shapes,arrows}

\title{Hybrid Renormalization for Quasi Distribution Amplitudes of A Light Baryon}

\author[a]{Chao Han}
\author[b,*]{Yushan Su}
\author[a,c]{Wei Wang} 
\author[a,*]{Jia-Lu Zhang\blfootnote{*Corresponding author}}

\affiliation[a]{INPAC, Key Laboratory for Particle Astrophysics and Cosmology (MOE), 
Shanghai Key Laboratory for Particle Physics and Cosmology, School of Physics and Astronomy, Shanghai Jiao Tong University, Shanghai 200240, China}
\affiliation[b]{Department of Physics, University of Maryland, College Park, MD 20742}
\affiliation[c]{Southern Center for Nuclear-Science Theory (SCNT), Institute of Modern Physics, Chinese Academy of Sciences, Huizhou 516000, Guangdong Province, China}

\emailAdd{chaohan@sjtu.edu.cn}
\emailAdd{ysu12345@umd.edu}
\emailAdd{wei.wang@sjtu.edu.cn}
\emailAdd{elpsycongr00@sjtu.edu.cn}

\abstract{
We develop a hybrid scheme to renormalize quasi distribution amplitudes of a light baryon on the lattice, which combines the self-renormalization and ratio scheme. 
By employing self-renormalization, the UV divergences and linear divergence at large spatial separations in quasi  distribution amplitudes are removed without introducing extra nonperturbative effects, while making a ratio with respect to the zero-momentum matrix element can properly remove the UV divergences in small spatial separations.
As a specific application,  distribution amplitudes  of the $\Lambda$ baryon made of $uds$ are investigated, and 
the requisite equal-time correlators, which define quasi distribution amplitudes in coordinate space, are perturbatively calculated up to the next-to-leading order in strong coupling constant $\alpha_s$. These perturbative equal-time correlators are used to convert lattice QCD matrix elements to the continuum space during the renormalization process. Subsequently, quasi distribution amplitudes are matched onto lightcone distribution amplitudes by integrating out hard modes and the corresponding hard kernels are derived up to next-to-leading order in $\alpha_s$ including the hybrid counterterms. 
These results are valuable in the lattice-based investigation of the lightcone distribution amplitudes of a light baryon from the first principles of QCD.
}
\date{\today}

\begin{document}

\maketitle
\flushbottom
\section{Introduction}

Lightcone distribution amplitudes (LCDAs) of light baryons are the fundamental non-perturbative inputs in QCD factorization for exclusive processes with a large momentum transfer~\cite{Shih:1998pb}. 
An example of this type is weak decays of bottom baryons which are valuable to extract the CKM matrix element  $|V_{ub}|$~\cite{LHCb:2015eia} and to probe new physics beyond the standard model through flavo\textit{}r changing neutral current process~\cite{LHCb:2015tgy,LHCb:2021byf}. 
In addition, the knowledge of LCDAs is also crucial for understanding the internal structure of baryons.  The LCDAs characterize the distributions of the longitudinal momentum among the quarks and gluons within the dominant leading Fock state of a baryon, which are complementary to parton distribution functions that encode the probability density distribution of parton momenta in hadrons. 

Many progresses have been made in exploring LCDAs of a nucleon with the  theoretical techniques such as  QCD sum rules~\cite{Chernyak:1984bm,King:1986wi,Braun:1999te,Anikin:2013aka} and Lattice QCD~\cite{Gockeler:2008xv,QCDSF:2008qtn,Bali:2015ykx,RQCD:2019hps} , but most of the available analyses are limited to the few lowest moments of LCDAs. Due to the lack of a complete knowledge of baryon LCDAs, many phenomenological analyses adopt model paramterizations resulting in uncontrollable errors in theoretical predictions for decay branching fractions of heavy baryons. Thus, it is highly indispensable to develop a method to calculate the full shape of baryon LCDAs from the first principle of QCD.

Compared to parton distribution functions (PDFs), 
LCDAs of light mesons and baryons  are more dificult to obtain. A major difficulty is that in an exclusive process it is very likely that more than one LCDAs enter into one physical observable in a rather complicated way through convolution integrals. Thus it makes an experimental determination of LCDAs extremely difficult. In addition, the leading-twist baryon LCDA $\Phi\left(y_1, y_2, \mu\right)$ with $y_1, y_2$ being the momentum fractions of two involved quarks and the momentum fraction for the third quark satisfying $y_3=1-y_1-y_2$  describes the momentum distributions of the three quarks, and by definition is a two-dimensional distribution function, which is even more complicated than the meson LCDAs. Our limited knowledge of baryon LCDAs mostly relies on non-perturbative methods, many of which are model-dependent and inevitably introduce uncontrollable uncertainties. In recent years lattice QCD  has been applied to determine the normalization constants and the ﬁrst moments of the distribution amplitudes for the lowest-lying baryon octet~\cite{QCDSF:2008qtn,Bali:2015ykx,RQCD:2019hps}. In particular, a latest lattice QCD study has used a large number of $n_f=2+1$ ensembles with physical pion (and kaon) masses and ﬁve diﬀerent lattice spacings~\cite{RQCD:2019hps}. After making the extrapolation to the continuum and inﬁnite volume limits, they obtained results for the first two moments of LCDAs. Despite these progresses, a complete description of LCDAs can not be constructed from these few moments and the LCDA is far from being deciphered.

In a previous publication~\cite{Deng:2023csv}, a direct method to extract the shape distribution of LCDA of a light baryon is proposed through the simulation of equal-time correlation functions, named as quasi distribution amplitude (DA),  under the framework of large momentum effective theory (LaMET)~\cite{Ji:2013dva,Ji:2014gla} (please see Refs.~\cite{Cichy:2018mum,Zhao:2018fyu,Ji:2020ect} for reviews on the development and successful applications in LaMET). 
The quasi-DA $\tilde \Phi (x_1,x_2,P^z,\mu)$ with $P^z$ being the hadron momentum on the $z$ direction and $x_1,x_2$ being the momentum fractions is a calculable quantity on the lattice. 
Since the quasi-DA and LCDA have the same infrared (IR) structure,  the LCDA of a light  baryon can be obtained by performing a boost on the hadron momentum to infinity, which is captured by the matching formula
\begin{equation}\label{matching}
\tilde{\Phi}\left(x_1, x_2, P^z,\mu\right)= \int d y_1 d y_2 \mathcal{C}\left(x_1, x_2, y_1, y_2, P^z,\mu\right) \Phi\left(y_1, y_2, \mu\right) +\mathcal{O}\left(\frac{\Lambda_{\rm QCD}}{x_1 P^z}, \frac{\Lambda_{\rm QCD}}{x_2 P^z}, \frac{\Lambda_{\rm QCD}}{\left(1-x_1-x_2\right) P^z}\right),
\end{equation}
where $\mathcal{C}(x_1,x_2,y_1,y_2,P^z,\mu)$ is a hard kernel to compensate for the ultraviolet (UV) differences between these two distributions. The hard kernel has been calculated up to one-loop accuracy in the $\overline {\rm MS}$ scheme~\cite{Deng:2023csv}.  
The   $\mu$ in the quasi-DA comes from the renormalization of the logarithmic divergences, while the scale $\mu$ in the LCDA is the factorization scale to split the collinear and hard modes. Thus the hard kernel $\mathcal{C}$ contains both renormalization and factorization scales, which are chosen to be the same for convenience. 
To remove the remnant ultraviolet divergence in the quasi-DA, it is attempted to adopt a regularization invariant momentum subtraction method (RI/MOM)~\cite{Martinelli:1994ty} to renormalize the quasi-DA and the corresponding one-loop counterterm was obtained~\cite{Deng:2023csv}.

In addition to the difficulty in implementing the RI/MOM scheme on the lattice,  discrepancies emerge when using this scheme to renormalize the quasi-PDFs and quasi-DAs on the lattice. 
Residual linear divergence continues to manifest~\cite{Zhang:2020rsx} and additional uncontrollable infrared effects are unavoidably introduced~\cite{LatticePartonCollaborationLPC:2021xdx}.
Moreover, the discretization effects arising in the lattice calculations also need to be considered carefully.
Given all these difficulties, it remains a challenge to perform a practical lattice calculation for baryon quasi-DA.

In order to address these issues, we develop a hybrid renormalization method following the same spirit of Ref.~\cite{Ji:2020brr}. The concept of the hybrid renormalization scheme was first proposed in Ref.~\cite{Ji:2020brr} in which the quasi-parton distribution functions were considered. The physical motivation for proposing the hybrid scheme was trying to purely deal with the renormalization of ultraviolet effects, without introducing extra infrared effects. The quasi distributions calculated on lattice contain the UV divergences, which are supposed to be described by asymptotically-free QCD perturbation theory. However, in some of the renormalization methods such as ratio scheme, uncontrollable non-perturbative IR effects are introduced at large distance. To overcome this issue, the hybrid renormalization was proposed in Ref.~\cite{Ji:2020brr}. It has a wide range of applications~\cite{Hua:2020gnw,Gao:2021dbh,LatticeParton:2022zqc,Chou:2022drv,Hua:2022wop,Gao:2022iex,LatticeParton:2022xsd,Su:2022fiu,Ji:2022ezo,Gao:2022ytj,Gao:2022uhg,Ji:2022thb,Zhang:2023tnc,Holligan:2023rex,Zhang:2023bxs,Gao:2023lny} since its proposal.

In this work, the hybrid scheme for the quasi-DA of a light baryon is developed utilizing the perturbative coordinate-space quasi-DA calculations, in which divergences appearing in both long-distance and short-distance spatial separations can be eliminated properly. 
By taking a ratio with respect to the zero-momentum matrix element at short distances, one can eliminate the UV divergences in lattice matrix elements, and part of the discretization effects are also expected to be canceled. Correspondingly, the short distance UV logarithms in the $\overline{\rm MS}$ scheme can also be eliminated properly. Through the self-renormalization at large distances, one can eliminate the UV divergences without introducing additional uncontrollable non-perturbative effects. Since the baryon DA is a two-dimensional distribution, there are multi regions involving both short distance and large distance simultaneously, which will be treated separately in this scheme.
As a result, this method ensures that lattice matrix elements approach the continuum limit in a more appropriate manner
and allows for a realistic determination of LCDA.

The rest of this paper is organized as follows: In Sec.~\ref{sec:LCDA_quasi_DA}, we give a brief overview of the lightcone distribution amplitudes and quasi-DA of a light baryon.
The detailed calculations of the one-loop spatial correlation are presented in Appendix~\ref{sec:LBDA_one_loop}.
In Sec.~\ref{sec:hybrid}, the hybrid renormalization scheme is developed and the matching kernel is presented. Some detailed results are collected in Appendix~\ref{sec:Integrals} and \ref{sec:douplusfunc}.
A summary is provided in the last section.

\section{Lightcone distribution amplitudes and quasi distribution amplitudes for a light baryon}
\label{sec:LCDA_quasi_DA}

In this section,  we introduce the requisite notations and conventions required for subsequent discussions.  In particular, we will give the definition of LCDAs, and quasi-DAs and collect the results for the one-loop matching in the ${\overline {\rm MS}}$ scheme. 

\subsection{LCDAs}

We start with the LCDAs, which are defined as the hadron-to-vacuum matrix elements of non-local operators consisting of quarks and gluon which live on the light cone.
In the case of a light baryon, the three-quark matrix element can be constructed as~\cite{Braun:1999te}
\begin{equation}
\left\langle 0\left|\varepsilon^{i j k} u_\alpha^{i^{\prime}}\left(z_1 \right)U_{i^{\prime} i}\left(z_1 , z_0 \right) d_\beta^{j^{\prime}}\left(z_2 \right)
U_{j^{\prime} j}\left(z_2 , z_0 \right) 
s_\gamma^{k^{\prime}}\left(z_3 \right)
U_{k^{\prime} k}\left(z_3 , z_0 \right)
\right| \Lambda(P, \lambda)\right\rangle,
\end{equation}
where $\left.\left.\right | \Lambda(P, \lambda)\right\rangle$ stands for the $\Lambda$ baryon state with the momentum $P$,  $P^2=0$ and the helicity $\lambda$.
$\alpha$, $\beta$ and $\gamma$ are Dirac indices.
$i^{(\prime)}$, $j^{(\prime)}$ and $k^{(\prime)}$ denote color charges.
In this paper, two light-cone unit vectors are defined as $n^\mu=(1,0,0,-1)/\sqrt{2}$ and $\bar n^\mu=(1,0,0,1)/\sqrt{2}$. 
The momentum of the baryon is along the $\bar n$ direction, $P^\mu=P^{+} \bar n^\mu = (P^z,0,0,P^z)$.
The coordinates are set in the $n$ direction, $z_i^\mu=z_i n^\mu$.  
The Wilson lines $U(x,y)$
\begin{equation}
U(x, y)=\mathcal{P} \exp \left[i g \int_0^1 \mathrm{~d} t(x-y)_\mu A^\mu(t x+(1-t) y)\right]
\end{equation}
are inserted to preserve the gauge invariance. In the definition, $z_0$ can be chosen freely due to the gauge invariance, and in the following we will use $z_0=0$ for simplicity.
Also for brevity Wilson lines, color indexes, and helicity will not be written out explicitly below.

Based on Lorentz invariance,  and the spin and parity requirement, the matrix element can be decomposed in terms of three functions, $V(z_i P\cdot n)$, $A(z_i P\cdot n)$, and $T(z_i P\cdot n)$ to the leading twist (twist-3)
\begin{align}
& \left\langle 0\left|u_\alpha^{}\left(z_1\right) d_\beta^{}\left(z_2\right) s_\gamma^{}\left(z_3\right) \right| \Lambda(P)\right\rangle 
\\
& ={f_N}\left\{( P\!\!\!\!/ C)_{\alpha \beta}\left(\gamma_5 u_{\Lambda}\right)_\gamma V\left(z_i P\cdot n\right)+\left( P\!\!\!\!/ \gamma_5 C\right)_{\alpha \beta} (u_{\Lambda})_\gamma A\left(z_i P\cdot n\right)
+\left(i \sigma_{\mu \nu} P^\nu C\right)_{\alpha \beta}\left(\gamma_\mu \gamma_5 u_{\Lambda}\right)_\gamma T\left(z_i P\cdot n\right)\right\} ,\notag
\end{align}
where C signifies the charge conjugation.
$u_{\Lambda}$ stands for the $\Lambda$ baryon spinor. Equivalently,  the three leading twist functions can be projected by inserting a specific gamma matrix $\Gamma$ into the $u$ and $d$ quark fields. 
In the following discussion, we will take $A(z_i P \cdot n)$ as an example while the other matrix elements can be similarly analyzed. 
Then we have 
\begin{equation}\label{eq:LCDA}
    \begin{aligned}
& M_L(z_1,z_2,z_3,P^+,\mu)=\left\langle 0\left|u^T\left(z_1\right) {\Gamma} d\left(z_2\right) s\left(z_3\right)\right| \Lambda(P)\right\rangle_R,
\\
&{\Phi_L}\left(x_1, x_2, \mu \right) f_{\Lambda}(\mu) P^{+} u_{\Lambda}(P)=\int_{-\infty}^{+\infty} \frac{d \, P^+ z_1}{2 \pi} \frac{d \, P^+ z_2}{2 \pi} e^{i x_1 P^+ z_1+i x_2 P^+ z_2} M_L(z_1,z_2,0,P^+,\mu),
\end{aligned}
\end{equation}
where $T$ means transpose and ${\Gamma}=C \gamma_5 \slashed n $.  
$R$ stands for renormalization. $x_i$s label the longitudinal momentum fractions carried by the three quarks and $0\leq x_i \leq 1$. The $\mu$ denotes the renormalization scale which will be converted to the factorization scale when the factorization of quasi-DA is established. $f_\Lambda(\mu)$ is the $\Lambda$ baryon decay constant defined as follows
\begin{equation}
f_\Lambda(\mu) P^{+} u_{\Lambda}(P) = M_L(0,0,0,P^+,\mu).
\end{equation}

It should be noted that we have defined the LCDA ${\Phi_L}\left(x_1, x_2, \mu \right)$ by separating the baryon decay constant $f_\Lambda(\mu)$, which has a different convention with the recent LQCD calculation~\cite{RQCD:2019hps}. 
Note that $f_\Lambda(\mu)$ depends on the renormalization scale $\mu$ since the local operator here is not a conserved current.   The LCDA ${\Phi_L}\left(x_1, x_2, \mu \right)$ in Eq.~(\ref{eq:LCDA}) is dimensionless and normalized.   


\subsection{Quasi-DAs}

\begin{figure}[htb]
\centering
\includegraphics[width=0.618\textwidth]{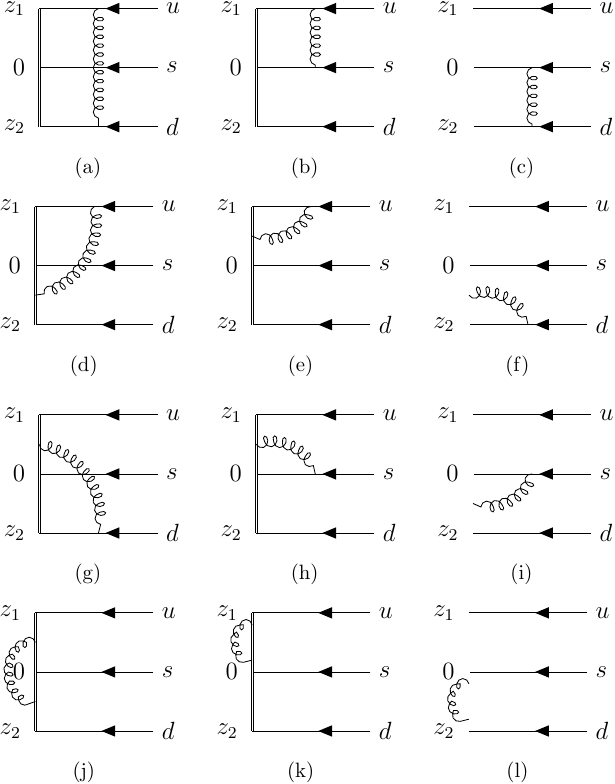}
\caption{One loop corrections for the equal-time matrix element of the $\Lambda$ baryon.}
\label{Pspic}
\end{figure}   


We consider a spatial correlator     
\begin{equation} \label{eq:MDA}
        M(z_1,z_2,z_3,P^{z},\mu)=\left\langle 0\left|u^T\left(z_1\right) \widetilde{\Gamma} d\left(z_2\right) s\left(z_3\right)\right| \Lambda(P)\right\rangle_R
\end{equation}
to define the quasi-DA
\begin{equation}
        \tilde{\Phi}\left(x_1, x_2, P^z,\mu\right) \tilde f_{\Lambda}(\mu) P^z u_{\Lambda}(P)=(P^z)^2\int_{-\infty}^{+\infty} \frac{d \, z_1}{2 \pi} \frac{d \, z_2}{2 \pi} e^{i x_1 P^z z_1+i x_2 P^z z_2} M(z_1,z_2,0,P^z,\mu),
\end{equation}
where $\tilde \Gamma= C \gamma_5 n\!\!\!\slash_z$.
For the quasi-DAs, the coordinates are set as $z_i^\mu=z_i n_z^\mu $, where $n_z^\mu=(0,0,0,1)$.
The baryon decay constant $\tilde  f_\Lambda(\mu)$ here can be similarly defined as Eq.~(\ref{eq:LCDA}):
\begin{eqnarray}
\tilde  f_\Lambda(\mu)=\frac{M(0,0,0,P^z,\mu)}{P^{z} u_{\Lambda}(P)}.
\end{eqnarray}

To obtain the quasi-DAs from the lattice simulations,  one needs to consider their short and long-distance properties separately. From this viewpoint, it is advantageous to work in the coordinate space.  To explicitly demonstrate the ultraviolet and infrared structures, we perform a  perturbative calculation of quasi-DAs by sandwiching the operators between the vacuum state $\left \langle0|\right.$ and the lowest-order Fock state $\left.|uds\right \rangle$:
\begin{equation}
{M}_{p}(z_1,z_2,z_3=0,P^z,\mu) = \left\langle 0\left|u^T\left(z_1\right) \widetilde\Gamma d\left(z_2\right) s\left(0\right)\right| u(x_1 P)d(x_2 P)s(x_3 P)\right\rangle_R,
\end{equation}
where the lower index ``$p$" in $M_{p}$ denotes the perturbative calculation. The next-to-leading order Feynman diagrams are shown in Fig.~\ref{Pspic}, and the calculation details are collected in Appendix~\ref{sec:LBDA_one_loop}. 
The diagrams in Fig.~\ref{Pspic} are categorized by the different divergences that show up, which is equivalent to the manner of gluon attachment to quarks or Wilson lines. For the quark-gluon-quark pattern collected in the first line, only infrared (IR) divergences exist. For the Wislon-line self-energy type diagrams shown in the fourth line, there are only ultraviolet (UV) divergences. For the quark-gluon-Wilson-line exchange diagrams in the second and third lines, both UV and IR divergences appear in the perturbative result.

The final results for the spatial correlator up to one loop are given as
 \begin{align} \label{eq:Mpert}
 &{M}_{p}(z_1,z_2,0,P^z,\mu)=
  \left\{1 
 + \frac{\alpha _s C_F}{\pi }\left(
 \frac{1}{2}  L_1^\text{UV}
 +
 \frac{1}{2}  L_2^\text{UV}
 +
 \frac{1}{2}  L_{12}^\text{UV}
 +\frac{3}{2}
 \right)\right\} 
 {M}_0 \left(z _1 , z _2 , 0 ,P^z,\mu\right) 
 \notag\\& 
-\frac{\alpha _s C_F}{8 \pi } \int_0^1 d \eta _1 \int_0^{1-\eta _1} d \eta _2  
 \notag\\& 
\times \left\{
\left (L_1^\text{IR}-1+\frac{1}{\epsilon_{\mathrm{IR}}}\right )
{M}_0 \left(\left(1-\eta _1\right) z _1 , z _2 , \eta _2 z _1 , P^z,\mu \right) 
+
\left (L_2^\text{IR}-1+\frac{1}{\epsilon_{\mathrm{IR}}}\right )
{M}_0 \left(z _1 , \left(1-\eta _1\right) z _2 , \eta _2 z _2 , P^z,\mu \right)
\right.
\notag\\&
\left.+2 
\left (L_{12}^\text{IR}-3+\frac{1}{\epsilon_{\mathrm{IR}}}\right )
{M}_0 \left(\left(1-\eta _1\right) z _1+\eta _1 z _2 , \left(1-\eta _2\right) z _2+\eta _2 z _1 , 0 , P^z,\mu \right) \right\}
 \notag\\& 
-\frac{\alpha _s C_F}{4 \pi } \int_0^1 d \eta  
\times\left\{{M}_0 \left((1-\eta ) z _1+\eta z _2 , z_2 , 0 , P^z,\mu \right) 
\left\{
\left (L_{12}^\text{IR}+1+\frac{1}{\epsilon_{\mathrm{IR}}}\right )
\left(\frac{1-\eta }{\eta }\right)_+ 
+2 \left(\frac{\ln  \eta }{\eta }\right)_+ \right\}  \right.
\notag \\& 
+{M}_0 \left(z _1 , (1-\eta ) z _2+\eta z _1 , 0 , P^z,\mu \right) 
\left\{ 
\left (L_{12}^\text{IR}+1+\frac{1}{\epsilon_{\mathrm{IR}}}\right )
\left(\frac{1-\eta }{\eta }\right)_+
+2 \left(\frac{\ln  \eta}{\eta }\right)_+ 
\right\} 
 \notag\\& 
+{M}_0 \left((1-\eta ) z _1 , z _2 , 0 , P^z,\mu \right) 
\left\{
\left(
L_1^\text{IR}+1+\frac{1}{\epsilon_{\mathrm{IR}}}
\right)
\left(\frac{1-\eta }{\eta }\right)_+ 
+2 \left(\frac{\ln  \eta  }{\eta }\right)_+ \right\}   
\notag \\& 
+{M}_0 \left(z _1 , (1-\eta ) z _2 , 0 , P^z,\mu \right) 
\left\{
\left(L_2^\text{IR}+1+\frac{1}{\epsilon_{\mathrm{IR}}}
\right)
\left(\frac{1-\eta }{\eta }\right)_+ 
+2 \left(\frac{\ln  \eta  }{\eta }\right)_+ \right\}  
\notag\\& 
+{M}_0 \left(z _1 , z _2 , \eta  z _1 , P^z,\mu \right) 
\left\{ 
\left (L_1^\text{IR}+1+\frac{1}{\epsilon_{\mathrm{IR}}}\right )
\left(\frac{1-\eta }{\eta }\right)_+
+2 \left(\frac{\ln  \eta }{\eta }\right)_+\right\} 
 \notag\\ &
\left.+{M}_0 \left(z _1 , z _2 , \eta  z _2 , P^z,\mu \right) 
\left\{ 
\left (L_2^\text{IR}+1+\frac{1}{\epsilon_{\mathrm{IR}}}\right )
\left(\frac{1-\eta }{\eta }\right)_+
+2 \left(\frac{\ln  \eta }{\eta }\right)_+\right\}\right\},
 \end{align} 
where $M_0$ stands for tree-level matrix element:
\begin{equation}
{M}_{0}(z_1,z_2,0,P^z,\mu) = \sqrt{2} P^z e^{i x_1 P^z z_1+i x_2 P^z z_2} u_s(x_3 P) ,
\end{equation}
where $x_3=1-x_1-x_2$, and   $\left[u_u\left(x_1 P\right)\right]^T \tilde{\Gamma} u_d\left(x_2 P\right)=\frac{1}{2} \operatorname{tr}\left[ P\!\!\!\slash C \gamma^5 \tilde{\Gamma}\right]$ is employed.
$u_{u/d/s}(P)$ denotes the spinor of u, d, or s quark with momentum $P$.
The plus function is defined as 
\begin{eqnarray}
    \displaystyle\int_0^1 d u \left[ G(u) \right]_{+} F(u)=\displaystyle\int_0^1 d u G(u) [F(u)-F(0)],
\end{eqnarray}
and some abbreviations are used in the above:
\begin{eqnarray}
L_1^{\text{IR, UV}}=\ln \left(\displaystyle\frac{1}{4}\mu_{\text{IR, UV}} ^2 z_1^2 e^{2 \gamma_E }\right), \;\;\;
L_2^{\text{IR,UV}}=\ln \left(\displaystyle\frac{1}{4}\mu_{\text{IR,UV}} ^2 z_2^2 e^{2 \gamma_E }\right), \\
L_{12}^{\text{IR,UV}}=\ln\left(\displaystyle\frac{1}{4}\mu^2_{\text{IR,UV}}(z_1-z_2)^2 e^{2\gamma_E}\right). 
\end{eqnarray}
We have checked that these results are consistent with the calculation in the momentum space~\cite{Deng:2023csv}. 
Moreover, one can see the UV and IR behaviors clearly in the coordinate space, which is convenient for the renormalization scheme to be established below.

Furthermore, one can obtain the zero-momentum matrix element in the coordinate space by letting~$P^z=0$ and performing the normalization with the local matrix element
\begin{equation}\label{eq:mo}
\begin{aligned}
&\hat{M}_{p}\left(z _1 , z _2 , z_3=0 , P^z=0,\mu \right) = \frac{{M}_{p}(z_1,z_2,0,0,\mu)}{{M}_{p}(0,0,0,0,\mu)} =  1 + \frac{\alpha_s C_F}{2 \pi} \left[ \frac{7}{8} L_1^{\text{UV}} + \frac{7}{8} L_2^{\text{UV}} + \frac{3}{4} L_{12}^{\text{UV}} + 4 \right],
\end{aligned}
\end{equation}
where the local matrix element $\displaystyle{M}_{p}(0,0,0,0,\mu) = \left(1 - \frac{\alpha_s C_F}{4 \pi} \frac{1}{\epsilon_{\rm IR}}\right) {M}_0 \left(0 , 0 , 0 ,0,\mu\right) $, 
see Eq.~(\ref{eq:2local}). The perturbative zero momentum matrix element in Eq.~(\ref{eq:mo}) will be used in the hybrid renormalization method.

\subsection{Matching of quasi-DAs in the ${\overline {\rm MS}}$ scheme}

In the large $P^z$ limit,  the quasi and lightcone distribution amplitudes can be related through the QCD factorization. After separating the hard and collinear contributions,  one can factorize the quasi-DAs in terms of the LCDAs and a hard kernel which can be perturbatively calculated.  
In the $\overline{\text{MS}}$ scheme, the  one-loop hard kernel has been obtained~\cite{Deng:2023csv}
\begin{equation}\label{eq:MSbarM}
\begin{aligned}
\mathcal{C}_{\rm \overline{MS}}\left(x_1, x_2, y_1, y_2, P^z,\mu\right) & = \delta\left(x_1-y_1\right) \delta\left(x_2-y_2\right)
+\frac{\alpha_s C_F}{8 \pi} 
 \times\left[C_2\left(x_1, x_2, y_1, y_2,P^z,\mu\right) \delta \left(x_2-y_2\right)\right.\\
& \left.+C_3\left(x_1, x_2, y_1, y_2,P^z,\mu\right) \delta\left(x_3-y_3\right) 
 +\left\{x_1 \leftrightarrow x_2, y_1 \leftrightarrow y_2\right\}\right]_{\oplus},
\end{aligned}
\end{equation}
where $\oplus$ denotes a double plus function   defined as 
\begin{equation} \label{dpf}
{\left[g\left(x_1, x_2, y_1, y_2\right)\right]_{\oplus}=}   g\left(x_1, x_2, y_1, y_2\right) 
 -\delta\left(x_1-y_1\right) \delta\left(x_2-y_2\right) 
 \int d x_1' d x_2' g\left(x_1', x_2', y_1, y_2\right). 
\end{equation}
$x_3=1-x_1-x_2$, $y_3=1-y_1-y_2$, and
\begin{align}
& C_2\left(x_1, x_2, y_1, y_2,P^z,\mu\right)= \\
& \left\{\begin{array}{l}
\displaystyle\frac{\left(x_1+y_1\right)\left(x_3+y_3\right) \ln \frac{y_1-x_1}{-x_1}}{y_1\left(y_1-x_1\right) y_3}-\frac{x_3\left(x_1+y_1+2 y_3\right) \ln \frac{x_3}{-x_1}}{\left(y_1-x_1\right) y_3\left(y_1+y_3\right)}, x_1<0 
\\
\displaystyle\frac{\left(x_1-3 y_1-2 y_3\right) x_1}{y_1\left(x_3-y_3\right)\left(y_1+y_3\right)}-\frac{\left[\left(x_3-y_3\right)^2-2 x_3 y_1\right] \ln \frac{x_3-y_3}{x_3}}{y_1\left(x_3-y_3\right) y_3}+\frac{2 x_1 \ln \frac{4 x_1\left(x_3-y_3\right) P_z^2}{\mu^2}}{y_1\left(x_3-y_3\right)}+\frac{x_1 \ln \frac{4 x_1 x_3 P_z^2}{\mu^2}}{y_1\left(y_1+y_3\right)}, 0<x_1<y_1 
\\
\displaystyle\frac{\left(x_3-2 y_1-3 y_3\right) x_3}{y_3\left(x_1-y_1\right)\left(y_1+y_3\right)}-\frac{\left[\left(x_1-y_1\right)^2-2 x_1 y_3\right] \ln \frac{x_1-y_1}{x_1}}{\left(x_1-y_1\right) y_1 y_3}+\frac{2 x_3 \ln \frac{4 x_3\left(x_1-y_1\right) P_z^2}{\mu^2}}{\left(x_1-y_1\right) y_3}+\frac{x_3 \ln \frac{4 x_1 x_3 P_z^2}{\mu^2}}{y_3\left(y_1+y_3\right)}, y_1<x_1<y_1+y_3 
\\
\displaystyle\frac{\left(x_1+y_1\right)\left(x_3+y_3\right) \ln \frac{y_3-x_3}{-x_3}}{y_1 y_3\left(y_3-x_3\right)}-\frac{x_1\left(x_3+2 y_1+y_3\right) \ln \frac{x_1}{-x_3}}{y_1\left(y_3-x_3\right)\left(y_1+y_3\right)}, x_1>y_1+y_3.
\end{array}\right. 
\notag\\
& C_3\left(x_1, x_2, y_1, y_2,P^z,\mu\right)= \\
& \left\{\begin{array}{l}
\displaystyle\frac{\left(x_1 x_2+y_1 y_2\right) \ln \frac{x_2-y_2}{x_2}}{y_1\left(x_2-y_2\right) y_2}-\frac{x_1\left(x_2+y_1\right) \ln \frac{-x_1}{x_2}}{y_1\left(x_2-y_2\right)\left(y_1+y_2\right)}, x_1<0 
\\
\displaystyle\frac{1}{x_1-y_1}+\frac{2 x_1+x_2}{y_1\left(y_1+y_2\right)}+\frac{\left[\left(x_1+y_2\right) y_1-x_1^2\right] \ln \frac{x_2-y_2}{x_2}}{y_1\left(x_2-y_2\right) y_2}+\frac{x_1 \ln \frac{4 x_1\left(x_2-y_2\right) P_z^2}{\mu^2}}{y_1\left(x_2-y_2\right)}+\frac{x_1 \ln \frac{4 x_1 x_2 P_z^2}{\mu^2}}{y_1\left(y_1+y_2\right)}, 0<x_1<y_1 
\\
\displaystyle\frac{1}{x_2-y_2}+\frac{x_1+2 x_2}{y_2\left(y_1+y_2\right)}+\frac{\left[\left(x_2+y_1\right) y_2-x_2^2\right] \ln \frac{x_1-y_1}{x_1}}{\left(x_1-y_1\right) y_1 y_2}+\frac{x_2 \ln \frac{4 x_2\left(x_1-y_1\right) P_z^2}{\mu^2}}{\left(x_1-y_1\right) y_2}+\frac{x_2 \ln \frac{4 x_1 x_2 P_z^2}{\mu^2}}{y_2\left(y_1+y_2\right)}, y_1<x_1<y_1+y_2 \\
\displaystyle\frac{\left(x_1 x_2+y_1 y_2\right) \ln \frac{x_1-y_1}{x_1}}{y_1\left(x_1-y_1\right) y_2}-\frac{x_2\left(x_1+y_2\right) \ln \frac{-x_2}{x_1}}{y_2\left(x_1-y_1\right)\left(y_1+y_2\right)}, x_1>y_1+y_2
.
\end{array}\right. \notag
&
\end{align}

It should be noticed that the above matching of quasi-DAs in the ${\overline {\rm MS}}$ scheme is problematic which can be understood as follows. As shown in Eq.~\eqref{dpf}, the double plus function contains an integral over the momentum fractions $x_1'$ and $x_2'$ which arises from the so-called virtual corrections to DAs~\cite{Deng:2023csv}. In the $x_1', x_2'\to \infty$ limit the integrals have the asymptotic form $\int dx'/x'$ and are then divergent. This divergence corresponds to the logarithmic UV divergence and should be renormalized. However, in deriving the above matching kernel,  one has only calculated the so-called real diagrams to quasi-DAs and assumed that virtual contributions are the integration of real diagrams and thus combining them gives the double plus function.  Keeping the remnant  UV divergence leads to an improper definition of double plus function. 

In Ref.~\cite{Izubuchi:2018srq}, a properly defined plus function is introduced for quasi parton distribution functions through the subtraction at the infinite momentum, while in Ref.~\cite{Deng:2023csv}, the RI/MOM renormalization scheme has been employed to subtract the asymptotic contributions for the quasi-DAs so that the plus function becomes well-defined. However it has been shown that implementing this scheme in the analysis of PDFs introduces unavoidably additional nonperturbative effects~\cite{LatticePartonCollaborationLPC:2021xdx} and thus the final results contain uncontrollable systematic uncertainties.

\section{Hybrid renormalization scheme for baryon quasi-DA}
\label{sec:hybrid}

In this section,  we will present the hybrid scheme to address the problems in the matching as stated in the previous section. The aim of this scheme is to remove the UV divergences in quasi-DAs without introducing nonperturbative effects, and meanwhile, this scheme can be properly implemented on the lattice.  

Conceptually, the UV divergences correspond to the small-distance behavior, while nonperturbative contributions are due to the long-distance behavior. So from the viewpoint of separating these contributions, it is advantageous to work in the coordinate space. 
In coordinate space, the results of quasi-DAs and the hard kernel can be obtained by making a Fourier transformation of the results in momentum space. We have also directly calculated the perturbative contributions to the quasi-DAs in Appendix~\ref{sec:LBDA_one_loop}, and checked the consistency.

At short distances, e.g. $z_1 \rightarrow 0$, $z_2 \rightarrow 0$ or $|z_1 - z_2| \rightarrow 0$ in Eq.~(\ref{eq:Mpert}) with $z_3=0$, perturbative quasi-DAs behave as the logarithmic terms like $\ln(z_1^2)$, $\ln(z_2^2)$ and $\ln((z_1-z_2)^2)$.   
As stated in the above section, the $\ln z_i^2$ terms in the real diagram of the perturbative matrix elements prevent them from performing $z\to 0$.
This is also reflected in our virtual contribution definition in which corresponding logarithm UV divergence had been kept.
Subsequently, the double plus function is also not well-defined.
By dividing the perturbative matrix element by a proper zero momentum matrix element, which is called the ratio scheme~\cite{Orginos:2017kos,Radyushkin:2017cyf,Radyushkin:2017lvu}, all these issues arising in the perturbative matrix elements calculation can be resolved.
By construction, in the ratio scheme, the $\ln z_i^2$ terms in the numerator and the denominator are equal, then will be canceled.
The corresponding virtual contributions will no longer be divergent.
Then the double plus function can be guaranteed to be well-defined.
Since now the perturbative matrix element is fine for $z\to 0$, the corresponding lattice matrix elements can also be constructed and used for matching.
It should be noted that the ratio scheme is only available in the perturbative region.
If one performs it over the long-distance region, the IR structure may be changed.
These require that only zero momentum matrix elements in the short-distance region can be chosen as denominators and the renormalization for the matrix elements in the long-distance region needs additional consideration.


At large distances, e.g. $z_1 \sim 1/\Lambda_{\rm QCD}$, $z_2 \sim 1/\Lambda_{\rm QCD}$ or $|z_1 - z_2| \sim 1/\Lambda_{\rm QCD}$ in Eq.~(\ref{eq:MDA}), a proper renormalization scheme should not introduce extra non-perturbative effects.  
In this work, we use the self-renormalization scheme as advocated in Ref.~\cite{LatticePartonCollaborationLPC:2021xdx}.
In the self-renormalization scheme, a renormalization factor $Z_R$, which includes all typical divergences and discretion errors, is defined.
One can use this renormalization factor to convert the lattice matrix elements to continuous matrix elements without bringing in any non-perturbative effects.
Besides the UV divergence inherited by dimensional regularization, the linear divergence from the lattice simulation is eliminated as well.
By definition, the calculation of the renormalization factors requires the input of the UV divergences and some parameters.
Specifically, the UV divergences are fitted from zero momentum lattice matrix elements at small lattice spacings, and these parameters can be obtained through matching the renormalized lattice matrix element to the continuum perturbative matrix element in the perturbative region.  
As a result, the long-distance regions which merely involve non-perturbative scales in the renormalization can also be handled.

A subtlety in applying the hybrid renormalization scheme to quasi-DAs of a light baryon is that there are multi regions involving both perturbative and non-perturbative scales, e.g. ($z_1 \rightarrow 0$ and $z_2 \sim 1/\Lambda_{\rm QCD}$), ($z_2 \rightarrow 0$ and $z_1 \sim 1/\Lambda_{\rm QCD}$), or ($|z_1-z_2| \rightarrow 0$, $z_1 \sim 1/\Lambda_{\rm QCD}$ and $z_2 \sim 1/\Lambda_{\rm QCD}$), see the blue bands in Fig.~\ref{pic:Renorm} except for the one around $z_1 \sim -z_2$. However, if the logarithmic divergences regarding different scales ($z_1$, $z_2$ and $|z_1-z_2|$) can be factorized separately, the UV divergences regarding those scales can be multiplicatively renormalized separately. If so, the ratio scheme can be taken for the perturbative scale and the self renormalization can be performed for the non-perturbative scale. From the one-loop result in Eq.~(\ref{eq:mo}), one can see that the logarithmic divergences indeed factorize, which allows to perform the hybrid renormalization. An all-order hybrid renormalization requires that the three scales $z_1^2$, $z_2^2$ and $(z_1-z_2)^2$ in the zero-momentum matrix element can be factorized out up to all orders in perturbation theory
$$
\hat{M}_{p}\left(z _1, z _2, 0, 0, \mu \right) =  
\left(1 + \sum_{n=1}\sum_{m=0}^{n} \alpha_s^n a_{n,m} L_1^{m}\right)
\left(1 + \sum_{n=1}\sum_{m=0}^{n} \alpha_s^n b_{n,m} L_2^{m}\right)
\left(1 + \sum_{n=1}\sum_{m=0}^{n} \alpha_s^n c_{n,m} L_{12}^{m}\right),
$$
where $ L_1 = \ln \frac{z_1^2 \mu^2 e^{2 \gamma_E}}{4} $, $ L_2 = \ln \frac{z_2^2 \mu^2 e^{2 \gamma_E}}{4} $ and $ L_{12} = \ln \frac{(z_1-z_2)^2 \mu^2 e^{2 \gamma_E}}{4} $.
At higher orders, there may be mixing terms involving two or three logs such as $\sim L_1 L_2$ and $\sim L_1 L_2 L_{12}$. It is non-trivial to prove that they can be factorized out, which requires further detailed investigations. We only focus on the renormalization and matching up to the one loop accuracy in this work.

In this section, we will give the normalization and self renormalization in the beginning, which provide the building blocks for hybrid renormalization. Then we provide a method for hybrid renormalization. The matching kernel in the hybrid scheme is presented at the end.   

\subsection{Normalization}
One starts from a quasi-DA matrix element on lattice $M\left(z_1, z_2, z_3=0, P^z, a\right)$, which is defined in Eq.~(\ref{eq:MDA}).  
On the lattice, the matrix element is  regularized with the lattice spacing $a$.  To ensure the normalization, one has 
\begin{equation}\label{eq:NormM}
\hat{M}\left(z_1, z_2, 0, P^z, a\right)=\frac{M\left(z_1, z_2, 0, P^z, a\right)}{M\left(0, 0, 0, P^z, a\right)},
\end{equation}
where part of the discretization effects are eliminated in the lattice matrix elements through the normalization. 

\subsection{Self renormalization}

In this subsection, we discuss the extraction of the renormalization factor $Z_R(z_1,z_2,a,\mu)$ from the  zero momentum matrix element $\hat{M}\left(z_1, z_2, 0, 0, a\right)$ (Eq.~(\ref{eq:NormM}) if $P^z = 0$). This renormalization factor will be applied to the large distances in hybrid renormalization.

According to Ref.~\cite{LatticePartonCollaborationLPC:2021xdx}, the renormalization factor is an asymptotic expansion with respect to $a$, with both power dependence and logarithmic dependence
\begin{equation}\label{eq:ZRm0}
Z_{R}(z_1,z_2,a,\mu) = \exp\Big[\left(\frac{k}{a \ln[a \Lambda_{\rm QCD}]} 
- m_{0}\right) \tilde{z}+\frac{\gamma_0}{b_0} \ln \bigg[\frac{\ln [1 /(a \Lambda_{\rm QCD})]}{\ln [\mu / \Lambda_{\rm \overline{MS}}]}\bigg]+\ln \left[1+\frac{d}{\ln (a \Lambda_{\rm QCD})}\right] + f(z_1,z_2)a \Big]\ ,
\end{equation}
where $\displaystyle\left(\frac{k}{a \ln[a \Lambda_{\rm QCD}]} 
- m_{0}\right) \tilde{z}$ is the linear divergence~\cite{Chen:2016fxx,Ji:2017oey,Ishikawa:2017faj,Green:2017xeu, Ji:2020brr} and the mass renormalization parameter~\cite{Ji:1995tm,Beneke:1998ui,Bauer:2011ws,Bali:2013pla,Zhang:2023bxs}, $\displaystyle\frac{\gamma_0}{b_0} \ln \bigg[\frac{\ln [1 /(a \Lambda_{\rm QCD})]}{\ln [\mu / \Lambda_{\rm \overline{MS}}]}\bigg]+\ln \left[1+\frac{d}{\ln (a \Lambda_{\rm QCD})}\right]$ contains the log divergence.  $f(z_1,z_2)a$ (or $f(z_1,z_2)a^2$) is the discretization effect. $\tilde{z}$ is the effective length for the linear divergence, which is defined as follows
\begin{eqnarray}
\tilde{z} = \left\{
        \begin{array}{ll}
            |z_1-z_2|, & \quad z_1 z_2 < 0 \\
            {\rm max}\left(|z_1|,|z_2|\right), & \quad z_1 z_2 \geq 0.
        \end{array}
    \right.
\end{eqnarray}

\begin{figure}[htb]
    \centering
    \includegraphics[width=0.618\textwidth]{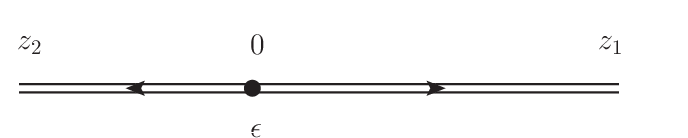}
    \caption{The schematic diagram of the Wilson links for $z_1 z_2 < 0$. $z_1$ and $z_2$ denote the locations of the heads of the Wilson links. The tails of the Wilson links are located at the origin, where the color indices are contracted with the Levi-Civita symbol $\epsilon$ to guarantee the gauge invariance.}
    \label{pic:efflenz1z2ll0}
\end{figure}  

\begin{figure}[htb]
    \centering
    \includegraphics[width=0.887\textwidth]{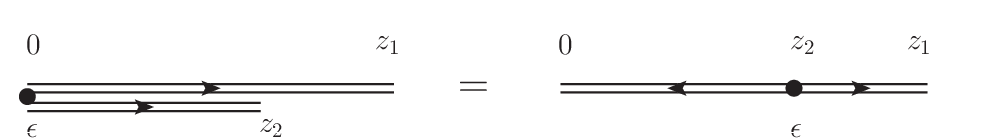}
    \caption{The schematic diagram of the Wilson links for $z_1 z_2 > 0$. The notations are the same as Fig.~\ref{pic:efflenz1z2ll0}. Considering the color contraction $\epsilon$, the two overlapped Wilson links in the same direction are equivalent to a Wilson link in the opposite direction. }
    \label{pic:efflenz1z2gg0}
\end{figure}

For $z_1 z_2 < 0$, the effective length is the total length of the Wilson link, see Fig.~\ref{pic:efflenz1z2ll0}. For $z_1 z_2 > 0$, there is an overlap region between the two Wilson links. However, one can simply show (e.g. $z_1 > z_2 > 0$)
\begin{eqnarray}
&U(z_1,0)_{i'i} U(z_2, 0)_{j'j} \epsilon_{ijk} 
= U(z_1,z_2)_{i'i''} U(z_2,0)_{i''i} U(z_2, 0)_{j'j} \epsilon_{ijk} \nonumber\\
&= U(z_1,z_2)_{i'i''} U(z_2,0)_{i''i} U(z_2, 0)_{j'j} \delta_{k k''}\epsilon_{ijk''}
\nonumber\\
&= U(z_1,z_2)_{i'i''} U(z_2,0)_{i''i} U(z_2, 0)_{j'j}  U^{\dag}(z_2,0)_{k k'} U(z_2,0)_{k' k''}\epsilon_{ijk''}\nonumber\\
&= U(z_1,z_2)_{i'i''} U^{\dag}(z_2,0)_{k k'} {\rm det}[U(z_2,0)] \epsilon_{i'' j' k'} 
= U(z_1,z_2)_{i'i''} U^{\dag}(z_2,0)_{k k'} \epsilon_{i''j'k'},
\end{eqnarray}
where the unitary and special properties of the Wilson links as SU(3) group elements are used in the second line and last line respectively. A schematic diagram for the above relation is shown in Fig.~\ref{pic:efflenz1z2gg0}. So the effective length is $z_1$ for $z_1 > z_2 > 0$. For a general case of $z_1 z_2 \geq 0$, the effective length should be the maximum value between $|z_1|$ and $|z_2|$. 

The leading order QCD beta function $\displaystyle b_0=\frac{11 C_A - 2 n_f}{6\pi}$ which satisfies $\displaystyle\frac{d \alpha_s}{d \ln[\mu]} = - b_0 \alpha_s^2$. The perturbative zero momentum matrix element in $\overline{\rm MS}$ scheme Eq.~(\ref{eq:mo}) satisfies the following renormalization group equation
\begin{equation}\label{eq:RGM0}
\frac{d \ln[\hat{M}_{p}\left(z_1, z_2, 0, 0, \mu\right)]}{d \ln[\mu]} = \gamma,
\end{equation}
where $\gamma=\gamma_0 \alpha_s + ...$ The leading anomalous dimension $\displaystyle\gamma_0=\frac{C_F}{2\pi}\left(5-\frac{7}{4}\delta_{z_1,0}-\frac{7}{4}\delta_{z_2,0}-\frac{3}{2}\delta_{z_1-z_2,0}\right)$ is scheme independent which can be applied to the renormalization factor Eq.~(\ref{eq:ZRm0}) of lattice matrix elements. It involves the quark-link interaction as well as the evolution effect of the local operator. There are subtraction terms since the UV fluctuation is frozen on lattice for a distance to be zero. $\Lambda_{\rm \overline{MS}}$ is the RG invariant scale for the LO running coupling, which is 0.142 GeV for $n_f=3$, 0.119 GeV for $n_f=4$ and 0.087 GeV for $n_f=5$, determined based on the method in~\cite{Karbstein:2018mzo}.  

The parameters $k$, $\Lambda_{\rm QCD}$, $f(z_1,z_2)$, $m_0$ and $d$ are extracted through the fit. The fit procedure is~\cite{LatticePartonCollaborationLPC:2021xdx}:\\ 
1) fit the $a$ dependence in $\hat{M}\left(z_1, z_2, 0, 0, a\right)$ to extract the global parameters $k$ and $\Lambda_{\rm QCD}$ as well as the discretization effect $f(z_1,z_2)$:
\begin{align}
&\hat{M}\left(z_1, z_2, 0, 0, a\right) \notag
\\&
= \exp\Big[\frac{k}{a \ln[a \Lambda_{\rm QCD}]}\tilde{z}+g(z_1,z_2,d)+\frac{\gamma_0}{b_0} \ln \bigg[\frac{\ln [1 /(a \Lambda_{\rm QCD})]}{\ln [\mu / \Lambda_{\rm \overline{MS}}]}\bigg]+\ln \left[1+\frac{d}{\ln (a \Lambda_{\rm QCD})}\right] + f(z_1,z_2)a \Big]\ ,
\end{align}
where $g(z_1,z_2,d)$ contains the non-perturbative intrinsic $z_1,z_2$ dependences, which is also extracted through the fit. $g(z_1,z_2,d)$ depends on the choice of the global parameter $d$ during the fit;\\
2) extract $m_0$ and $d$ through requiring the renormalized matrix element to be equal to the perturbative matrix element at short distances ($a < z_1,z_2 \ll 1/\Lambda_{\rm QCD}$)
\begin{equation}
\frac{\hat{M}\left(z_1, z_2, 0, 0, a\right)}{Z_{R}(z_1,z_2,a,\mu)}=\exp\Big[ g(z_1,z_2,d) + m_0 \tilde{z} \Big]=\hat{M}_{p}\left(z_1, z_2, 0, 0, \mu\right).
\end{equation}
The perturbative zero momentum matrix element is obtained from Eqs.~(\ref{eq:Mpert}) and (\ref{eq:mo}),
\begin{equation}
\begin{aligned}
&\hat{M}_{p}\left(z_1, z_2, 0, 0, \mu\right) = 1 + \frac{\alpha_s C_F}{2 \pi} \left[ \frac{7}{8} \ln\left(\frac{z_1^2 \mu^2 e^{2 \gamma_E}}{4}\right) + \frac{7}{8} \ln\left(\frac{z_2^2 \mu^2 e^{2 \gamma_E}}{4}\right) + \frac{3}{4} \ln\left(\frac{(z_1-z_2)^2 \mu^2 e^{2 \gamma_E}}{4}\right) + 4 \right],
\end{aligned}
\end{equation}
where the IR poles are canceled through the normalization with the local matrix element. The details of the perturbative calculation are presented in Appendix~\ref{sec:LBDA_one_loop}.

Thus one can define the renormalized lattice matrix element in $\overline{\rm MS}$ scheme for the whole range as the ratio of the normalized lattice matrix element Eq.~(\ref{eq:NormM}) to the renormalization factor Eq.~(\ref{eq:ZRm0})
\begin{equation}\label{eq:latinMSbar}
\hat{M}_{\overline{\rm MS}}\left(z_1, z_2, 0, P^z, \mu\right) = \frac{\hat{M}\left(z_1, z_2, 0, P^z, a\right)}{Z_{R}(z_1,z_2,a,\mu)},
\end{equation}
where the renormalization factor $Z_{R}$, though extracted from the zero momentum matrix element, can be applied to the large momentum matrix element since the renormalization is independent of the external states.

\subsection{Hybrid renormalization}

A hybrid renormalization method is presented in this subsection, based on the building blocks provided in the previous subsections, such as the normalized lattice matrix element Eq.~(\ref{eq:NormM}), the renormalization factor Eq.~(\ref{eq:ZRm0}) and the renormalized lattice matrix element in $\overline{\rm MS}$ scheme Eq.~(\ref{eq:latinMSbar}).


\begin{figure}[htb]
    \centering
    \includegraphics[width=0.618\textwidth]{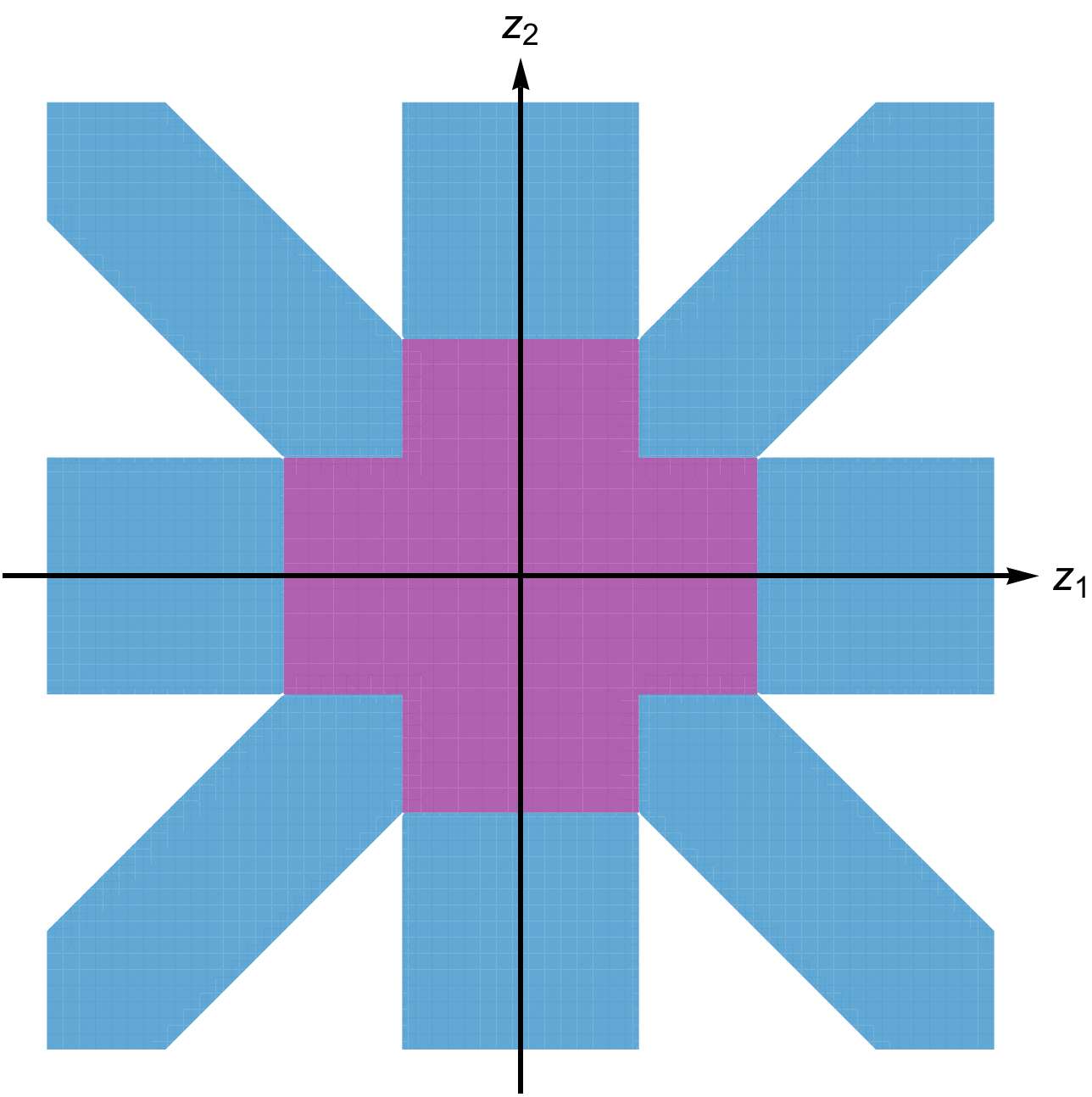}
    \caption{A schematic diagram of renormalization. First, both large and zero momentum matrix elements are converted to $\overline{\rm MS}$ scheme. Then, additional ratios are taken with the $\overline{\rm MS}$ scheme matrix elements as follows. The large momentum matrix element in the purple (hard) region is divided by the zero momentum matrix element in the purple region correspondingly. The large momentum matrix in the French blue (hard-soft) region is divided by the zero momentum matrix element on the blue-purple boundary. The large momentum matrix in the white (soft) region is divided by the zero momentum matrix element at the white-purple intersection point.}
    \label{pic:Renorm}
\end{figure}  

In practical calculations, it is difficult to directly perform the matching in $\overline{\rm MS}$ scheme since there are inconsistencies between the lattice matrix element and continuum scheme at short distances. 
The lattice matrix elements are finite as $z_1 \rightarrow 0$ or $z_2 \rightarrow 0$ while the perturbative matrix elements, with the logarithmic terms $\ln(z_1^2)$, $\ln(z_2^2)$ and $\ln((z_1-z_2)^2)$, are divergent as $z_1 \rightarrow 0$ or $z_2 \rightarrow 0$. 
Moreover, those logarithmic terms correspond to slowly decaying terms $\sim \displaystyle\frac{1}{|x'_1|}$ and $\displaystyle\frac{1}{|x'_2|}$ in the matching kernel, which increase the difficulties of numerical calculations in preserving the normalization, though not impossible. The hybrid scheme can be viewed as a modification of $\overline{\rm MS}$ scheme at short distance. Through the ratio at short distance, part of the discretization effects are canceled in the lattice matrix elements and the singular log terms are canceled in the perturbative matrix elements. So the lattice matrix elements become more consistent with the continuum scheme under the hybrid scheme, where it is easier to preserve the normalization.

Thus there are several principles in designing the hybrid scheme~\cite{LatticePartonCollaborationLPC:2021xdx}: \\
1) Eliminate all the singular logarithmic terms in the perturbative matrix elements including $\ln(z_1^2)$, $\ln(z_2^2)$ and $\ln((z_1-z_2)^2)$ through the ratio; \\
2) Avoid introducing extra effects that are not perturbatively controllable at large distances; \\
3) Keep the renormalized matrix element continuous and the method as simple as possible.

A hybrid renormalization method that satisfies the above principles is presented in the following, and a schematic diagram is shown in Fig.~\ref{pic:Renorm}.
For the sake of convenience, besides $|z_1|$, $|z_2|$ and $|z_1-z_2|$, we also treat $|z_1+z_2|$ as an argument.
The purple region denotes that all arguments are perturbative scales.
The white regions denote that all are non-perturbative scales.
The blue regions denote that only one of the four scales  is perturbative.
\begin{itemize}
\item If the scales $z_1$, $z_2$ and $z_1-z_2$ are all perturbative, e.g. ($|z_1|<z_s$ and $|z_2|<z_s$) or ($|z_1|<z_s$ and $z_s<|z_2|<2z_s$) or ($z_s<|z_1|<2z_s$ and $|z_2|<z_s$), which corresponds to the purple region in Fig.~\ref{pic:Renorm}, one introduces the ratio scheme on the normalized lattice matrix elements (Eq.~(\ref{eq:NormM}))
 \begin{equation}
 \begin{aligned}
&\frac{\hat{M}\left(z_1, z_2, 0, P^z, a\right)}{\hat{M}\left(z_1, z_2, 0, 0,a\right)} \left(\theta(z_s-|z_1|)\theta(z_s-|z_2|) +\theta(z_s-|z_2|)\theta(|z_1|-z_s)\theta(2z_s-|z_1|) \right.\\
&\left.+\theta(z_s-|z_1|)\theta(|z_2|-z_s)\theta(2z_s-|z_2|)\right) \\
&=\frac{\hat{M}_{\overline{\rm MS}}\left(z_1, z_2, 0, P^z,\mu\right)}{\hat{M}_{\overline{\rm MS}}\left(z_1, z_2, 0, 0,\mu\right)} \left(\theta(2z_s-|z_1|)\theta(z_s-|z_2|)+\theta(z_s-|z_1|)\theta(|z_2|-z_s)\theta(2z_s-|z_2|)\right),
\end{aligned}
\end{equation}  
where $z_s$ is the hybrid cutoff, which satisfies $a \ll 2z_s \ll 1/\Lambda_{\rm QCD}$. So no extra non-perturbative effects are introduced. The ratio can be written with the renormalized lattice matrix element in $\overline{\rm MS}$ scheme Eq.~(\ref{eq:latinMSbar}) since the renormalization factor $Z_R$ is independent of momentum $P^z$. There are several advantages to taking the ratio scheme at short distances. First, part of the discretization effects are canceled since $a \ll z_s$. Second, the normalization can be guaranteed in both renormalized lattice matrix elements and perturbative matching. In the perturbative matrix element, we will see the cancellation of $\ln(z_1^2)$, $\ln(z_2^2)$ and $\ln((z_1-z_2)^2)$ through the ratio, which leads to a current conserved matching kernel after Fourier transformation.

\item If $z_1$ is perturbative while $z_2$ and $z_1-z_2$ are not, e.g. $|z_1|<z_s$ and $|z_2|>2z_s$, which is the blue vertical region in Fig.~\ref{pic:Renorm}, one needs to introduce the ratio scheme for $z_1$ and $\overline{\rm MS}$ scheme (Eq.~(\ref{eq:latinMSbar})) for $z_2$,
\begin{equation}
\begin{aligned}
&\frac{\hat{M}\left(z_1, z_2, 0, P^z, a\right) Z_R(z_1,{\rm sign}(z_2)2z_s,a,\mu)}{\hat{M}\left(z_1, {\rm sign}(z_2)2z_s, 0, 0, a\right) Z_R(z_1,z_2,a,\mu)}\theta(z_s-|z_1|)\theta(|z_2|-2z_s) \\
&=\frac{\hat{M}_{\overline{\rm MS}}\left(z_1, z_2, 0, P^z, \mu\right)}{\hat{M}_{\overline{\rm MS}}\left(z_1, {\rm sign}(z_2)2z_s, 0, 0, \mu\right)}\theta(z_s-|z_1|)\theta(|z_2|-2z_s),
\end{aligned}
\end{equation}
where $Z_R(z_1,z_2,a,\mu)$ is the renormalization factor to convert the lattice matrix element to $\overline{\rm MS}$ scheme as we have defined in Eq.~(\ref{eq:ZRm0}). The zero momentum matrix element in the denominator $\hat{M}\left(z_1, {\rm sign}(z_2)2z_s, 0, 0, a\right)$ will be crucial in canceling the $\ln(z_1^2)$ in the perturbative matrix element when we deduce the matching kernel. 
No extra non-perturbative effect is introduced in the denominator where the $z_2$ dependence is truncated at ${\rm sign}(z_2)2z_s$.

\item If $z_2$ is perturbative while $z_1$ and $z_1-z_2$ are not, e.g. $|z_1|>2z_s$ and $|z_2|<z_s$, which is the blue horizontal region in Fig.~\ref{pic:Renorm}, one follows the similar strategy, 
\begin{equation}
\frac{\hat{M}_{\overline{\rm MS}}\left(z_1, z_2, 0, P^z, \mu\right)}{\hat{M}_{\overline{\rm MS}}\left({\rm sign}(z_1)2z_s, z_2, 0, 0, \mu\right)}\theta(|z_1|-2z_s)\theta(z_s-|z_2|).
\end{equation}

\item If both $z_1$ and $z_2$ are non-perturbative while $z_1-z_2$ is perturbative, e.g. $|z_1|>z_s$, $|z_2|>z_s$ and $|z_1-z_2|<z_s$,  which is the blue diagonal region around $z_1 \sim z_2$ in Fig.~\ref{pic:Renorm}, one takes
\begin{equation}
\frac{\hat{M}_{\overline{\rm MS}}\left(z_1, z_2, 0, P^z, \mu\right)}{\hat{M}_{\overline{\rm MS}}\left(z_1^*, z_2^*, 0, 0, \mu\right)}\theta(|z_1|-z_s)\theta(|z_2|-z_s)\theta(z_s-|z_1-z_2|),
\end{equation}
where $z_1^*=z_s+(z_1-z_2)\theta(z_1-z_2)$ and $z_2^*=z_s+(z_2-z_1)\theta(z_2-z_1)$. The coordinate choices $z_1^*$ and  $z_2^*$ apply for $z_1<0$ and $z_2<0$ as well because ($z_1 \rightarrow -z_2, z_2 \rightarrow -z_1$) is a symmetry for zero momentum matrix element. The zero momentum matrix element in the denominator will be crucial in canceling the $\ln((z_1-z_2)^2)$ in the perturbative matrix element. 

\item For $|z_1|>z_s$, $|z_2|>z_s$ and $|z_1 + z_2|<z_s$, which is the blue diagonal region around $z_1 \sim -z_2$ in Fig.~\ref{pic:Renorm}, one can introduce the ratio as well
\begin{equation}
\frac{\hat{M}_{\overline{\rm MS}}\left(z_1, z_2, 0, P^z, \mu\right)}{\hat{M}_{\overline{\rm MS}}\left(z_1^{**}, z_2^{**}, 0, 0, \mu\right)}\theta(|z_1|-z_s)\theta(|z_2|-z_s)\theta(z_s-|z_1+z_2|),
\end{equation}
where $z_1^{**}=z_s+(z_1+z_2)\theta(z_1+z_2)$ and $z_2^{**}=-z_s+(z_2+z_1)\theta(-z_2-z_1)$. This step is for continuity and simplicity. 

\item Finally, for $|z_1|>z_s$, $|z_2|>z_s$, $|z_1-z_2|>z_s$ and $|z_1+z_2|>z_s$, we apply the $\overline{\rm MS}$ scheme (Eq.~(\ref{eq:latinMSbar})),
\begin{equation}
\begin{aligned}
&\frac{\hat{M}\left(z_1, z_2, 0, P^z, a\right)}{Z_R(z_1,z_2,a,\mu) \hat{M}_{\overline{\rm MS}}\left({\rm sign}(z_1)z_s, {\rm sign}(z_2)2z_s, 0, 0, \mu\right)}\theta(|z_1|-z_s)\theta(|z_2|-z_s)\theta(|z_1-z_2|-z_s)\theta(|z_1+z_2|-z_s)\\
&=\frac{\hat{M}_{\overline{\rm MS}}\left(z_1, z_2, 0, P^z, \mu\right)}{\hat{M}_{\overline{\rm MS}}\left({\rm sign}(z_1)z_s, {\rm sign}(z_2)2z_s, 0, 0, \mu\right)}\theta(|z_1|-z_s)\theta(|z_2|-z_s)\theta(|z_1-z_2|-z_s)\theta(|z_1+z_2|-z_s).
\end{aligned}
\end{equation}
The isospin and parity symmetries allow different choices for the denominator: 
$$
\hat{M}_{\overline{\rm MS}}\left(z_s, 2z_s, 0, 0, \mu\right)=\hat{M}_{\overline{\rm MS}}\left(2z_s, z_s, 0, 0, \mu\right)=\hat{M}_{\overline{\rm MS}}\left(-z_s, -2z_s, 0, 0, \mu\right)=\hat{M}_{\overline{\rm MS}}\left(-2z_s, -z_s, 0, 0, \mu\right),
$$
$$\hat{M}_{\overline{\rm MS}}\left(-z_s, 2z_s, 0, 0, \mu\right)=\hat{M}_{\overline{\rm MS}}\left(2z_s, -z_s, 0, 0, \mu\right)=\hat{M}_{\overline{\rm MS}}\left(z_s, -2z_s, 0, 0, \mu\right)=\hat{M}_{\overline{\rm MS}}\left(-2z_s, z_s, 0, 0, \mu\right).$$
\end{itemize}

To conclude, the hybrid renormalized matrix element is
\begin{equation}\label{eq:Mhybrid}
\begin{aligned}
&\hat{M}_{H}(z_1,z_2,0,P^z) = \frac{\hat{M}_{\overline{\rm MS}}\left(z_1, z_2, 0, P^z,\mu\right)}{\hat{M}_{\overline{\rm MS}}\left(z_1, z_2, 0, 0,\mu\right)} \left(\theta(2z_s-|z_1|)\theta(z_s-|z_2|)+\theta(z_s-|z_1|)\theta(|z_2|-z_s)\theta(2z_s-|z_2|)\right) \\
&+\frac{\hat{M}_{\overline{\rm MS}}\left(z_1, z_2, 0, P^z, \mu\right)}{\hat{M}_{\overline{\rm MS}}\left(z_1, {\rm sign}(z_2)2z_s, 0, 0, \mu\right)}\theta(z_s-|z_1|)\theta(|z_2|-2z_s)+\frac{\hat{M}_{\overline{\rm MS}}\left(z_1, z_2, 0, P^z, \mu\right)}{\hat{M}_{\overline{\rm MS}}\left({\rm sign}(z_1)2z_s, z_2, 0, 0, \mu\right)}\theta(|z_1|-2z_s)\theta(z_s-|z_2|) \\
&+\frac{\hat{M}_{\overline{\rm MS}}\left(z_1, z_2, 0, P^z, \mu\right)}{\hat{M}_{\overline{\rm MS}}\left(z_s+(z_1-z_2)\theta(z_1-z_2), z_s+(z_2-z_1)\theta(z_2-z_1), 0, 0, \mu\right)}\theta(|z_1|-z_s)\theta(|z_2|-z_s)\theta(z_s-|z_1-z_2|) \\
&+ \frac{\hat{M}_{\overline{\rm MS}}\left(z_1, z_2, 0, P^z, \mu\right)}{\hat{M}_{\overline{\rm MS}}\left(z_s+(z_1+z_2)\theta(z_1+z_2), -z_s+(z_2+z_1)\theta(-z_2-z_1), 0, 0, \mu\right)}\theta(|z_1|-z_s)\theta(|z_2|-z_s)\theta(z_s-|z_1+z_2|) \\
&+\frac{\hat{M}_{\overline{\rm MS}}\left(z_1, z_2, 0, P^z, \mu\right)}{\hat{M}_{\overline{\rm MS}}\left({\rm sign}(z_1)z_s, {\rm sign}(z_2)2z_s, 0, 0, \mu\right)}\theta(|z_1|-z_s)\theta(|z_2|-z_s)\theta(|z_1-z_2|-z_s)\theta(|z_1+z_2|-z_s),
\end{aligned}
\end{equation}
where $\hat{M}_{\overline{\rm MS}}$ is the renormalized lattice matrix element in $\overline{\rm MS}$ scheme defined in Eq.~(\ref{eq:latinMSbar}). 

Before closing this subsection, we point out that in the hybrid renormalization the two schemes are bridged at $z=z_s$:  for $z_i < z_s$, we use the ratio scheme, while for $z_i > z_s$, the self renormalization is adopted. On the one hand the $z_s$ should be in principle larger than the lattice spacing $a$ so that discretization effects can be suppressed through the ratio at short distance. On the other hand, $z_s$ should be smaller than $1/\Lambda_{\rm QCD}$ so that no extra non-perturbative effects are introduced at large distance. To conclude, $z_s$ is the point where the continuum perturbation theory matches lattice data.

Explicitly the $z_s$ might be chosen around $0.1-0.2$ fm. For example, it is shown in Figs.~11, 12-16 (d) of Ref.~\cite{LatticePartonCollaborationLPC:2021xdx} and Fig.~1 of  Ref.~\cite{LatticeParton:2022zqc} that the continuum perturbation theory can describe lattice matrix element around $0.1-0.2$ fm. This value is much smaller than the proton charge radius around $0.8$ fm where the non-perturbative effects are important. Also this  corresponds to the energy scale $1-2$ GeV, which is typically larger than the non-perturbative scale $\Lambda_{\rm QCD} \sim 300$ MeV.

\subsection{Matching in hybrid scheme}
The hybrid scheme quasi baryon DA is obtained through the Fourier transformation
\begin{equation}
\tilde \Phi_{H} (x_1,x_2,P^z) = (P^z)^2 \int_{-\infty}^{+\infty} \frac{d z_1}{2\pi} \frac{d z_2}{2\pi} e^{i x_1 P^z z_1+ i x_2 P^z z_2} \hat{M}_{H}\left(z_1,z_2,0,P^z\right).
\end{equation}
The factorization formula is 
\begin{equation}
\tilde \Phi_{H} (x_1,x_2,P^z) = \int d y_1 d y_2 \mathcal{C}_{H}(x_1,x_2,y_1,y_2,P^z,\mu) \Phi_{L} (y_1,y_2,\mu),
\end{equation}
where $\Phi_L (y_1,y_2,\mu)$ is the baryon LCDA in Eq.~(\ref{eq:LCDA}). $\mathcal{C}_{H}(x_1,x_2,y_1,y_2,P^z,\mu)$ is the matching kernel in hybrid scheme, 
\begin{equation}
\mathcal{C}_{H}(x_1,x_2,y_1,y_2,P^z,\mu) = \mathcal{C}_{\overline{\rm MS}}(x_1,x_2,y_1,y_2,P^z,\mu) - \delta \mathcal{C}_{H}(x_1,x_2,y_1,y_2,P^z,\mu),
\end{equation}
where $\mathcal{C}_{\overline{\rm MS}}(x_1,x_2,y_1,y_2,P^z,\mu)$ is the $\overline{\rm MS}$ scheme matching kernel in Eq.~(\ref{eq:MSbarM}). $\delta \mathcal{C}_{H}(x_1,x_2,y_1,y_2,P^z,\mu)$ is the hybrid counterterm and the NLO result can be obtained through the Fourier transformation~\cite{Izubuchi:2018srq}
\begin{equation}
\begin{aligned}
&\delta \mathcal{C}_{H}^{(1)}(x_1,x_2,y_1,y_2,P^z,\mu) =(P^z)^2\int \frac{d z_1}{2\pi} \frac{d z_2}{2\pi} e^{i (x_1-y_1)P^z z_1 +  i (x_2-y_2)P^z z_2} \delta \hat{M}_{H}^{(1)}\left(z_1, z_2, \mu \right).
\end{aligned}
\end{equation}
$\delta \hat{M}_{H}^{(1)}\left(z_1, z_2, \mu \right)$ is the perturbative correction in the hybrid scheme at NLO:
\begin{align}\label{eq:Matching}
&\delta \hat{M}_{H}^{(1)}\left(z_1, z_2, \mu \right) = \hat{M}_{p}^{(1)}\left(z_1, z_2, 0, 0,\mu\right) \left(\theta(2z_s-|z_1|)\theta(z_s-|z_2|)+\theta(z_s-|z_1|)\theta(|z_2|-z_s)\theta(2z_s-|z_2|)\right) \notag\\
&+\hat{M}_{p}^{(1)}\left(z_1, {\rm sign}(z_2)2z_s, 0, 0, \mu\right)\theta(z_s-|z_1|)\theta(|z_2|-2z_s)+\hat{M}_{p}^{(1)}\left({\rm sign}(z_1)2z_s, z_2, 0, 0, \mu\right)\theta(|z_1|-2z_s)\theta(z_s-|z_2|) \notag\\
&+\hat{M}_{p}^{(1)}\left(z_s+(z_1-z_2)\theta(z_1-z_2), z_s+(z_2-z_1)\theta(z_2-z_1), 0, 0, \mu\right)\theta(|z_1|-z_s)\theta(|z_2|-z_s)\theta(z_s-|z_1-z_2|) \notag\\
&+ \hat{M}_{p}^{(1)}\left(z_s+(z_1+z_2)\theta(z_1+z_2), -z_s+(z_2+z_1)\theta(-z_2-z_1), 0, 0, \mu\right)\theta(|z_1|-z_s)\theta(|z_2|-z_s)\theta(z_s-|z_1+z_2|) \notag\\
&+\hat{M}_{p}^{(1)}\left({\rm sign}(z_1)z_s, {\rm sign}(z_2)2z_s, 0, 0, \mu\right)\theta(|z_1|-z_s)\theta(|z_2|-z_s)\theta(|z_1-z_2|-z_s)\theta(|z_1+z_2|-z_s).
\end{align}
$\hat{M}_{p}^{(1)}\left(z_1, z_2, 0, 0, \mu \right)$ is the NLO result of Eq.~(\ref{eq:mo}),
$$
\hat{M}_{p}^{(1)}\left(z_1, z_2, 0, 0, \mu \right) = \frac{\alpha_s C_F}{2 \pi} \left[ \frac{7}{8} \ln\left(\frac{z_1^2 \mu^2 e^{2 \gamma_E}}{4}\right) + \frac{7}{8} \ln\left(\frac{z_2^2 \mu^2 e^{2 \gamma_E}}{4}\right) + \frac{3}{4} \ln\left(\frac{(z_1-z_2)^2 \mu^2 e^{2 \gamma_E}}{4}\right) + 4 \right].
$$

One performs the Fourier transformation on different regions and obtains the hybrid counterterm:
\begin{align}\label{eq-f}
&\delta \mathcal{C}_{H}^{(1)}(x_1,x_2,y_1,y_2,P^z,\mu) = (P^z)^2 \frac{\alpha_s C_F}{2 \pi} \Bigg[ I_{\rm H} [(x_1-y_1)P^z,(x_2-y_2)P^z] + I_{\rm HSI} [(x_1-y_1)P^z,(x_2-y_2)P^z] \notag\\
&+ I_{\rm HSII} [(x_1-y_1)P^z,(x_2-y_2)P^z] + I_{\rm HSIII} [(x_1-y_1)P^z,(x_2-y_2)P^z] + I_{\rm HSIV} [(x_1-y_1)P^z,(x_2-y_2)P^z] \notag\\
&+ I_{\rm S} [(x_1-y_1)P^z,(x_2-y_2)P^z] + \delta[(x_1-y_1)P^z] \delta[(x_2-y_2)P^z] \left(\frac{5}{2} \ln\left(\frac{\mu^2 e^{2 \gamma_E}}{4}\right) + 4\right) \Bigg],
\end{align}
where all integrated formulas are collected in Appendix~\ref{sec:Integrals}.

Since the normalization with respect to the local matrix element is performed and the ratio is taken with respect to the zero momentum matrix element at short distances, the matching in the hybrid scheme preserves the normalization requirement.  
That means, if the light cone DA is normalized $\int dy_1 dy_2 \Phi_{L} (y_1,y_2,\mu)=1$, the quasi-DA obtained through matching in the hybrid scheme is also normalized $\int dx_1 dx_2 \tilde \Phi_{H} (x_1,x_2,P^z)=1$. As shown in Appendix~\ref{sec:douplusfunc}, if the normalization is preserved, the hybrid matching kernel at one loop is a pure double plus function defined as Eq.~(\ref{dpf}),
\begin{equation}\label{eq:dpfhybrid}
\mathcal{C}_{H}^{(1)}(x_1,x_2,y_1,y_2,P^z,\mu) = \left[\mathcal{C}^{(1)}_{\overline{\rm MS}}(x_1,x_2,y_1,y_2,P^z,\mu) - \delta \mathcal{C}^{(1)}_{H}(x_1,x_2,y_1,y_2,P^z,\mu)\right]_{\oplus}.
\end{equation}
A necessary condition for the pure double plus function is that there are no linearly decaying terms $\sim \displaystyle\frac{1}{|x'_1|}$ and $\displaystyle\frac{1}{|x'_2|}$ for $|x'_1|\gg 1$ and $|x'_2|\gg 1$ respectively. One can check those linearly decaying terms in $\mathcal{C}^{(1)}_{\overline{\rm MS}}(x_1,x_2,y_1,y_2,P^z,\mu)$ are explicitly cancelled by the linearly decaying terms in $\delta \mathcal{C}^{(1)}_{H}(x_1,x_2,y_1,y_2,P^z,\mu)$, which is equivalent to the cancellation of the log terms $\ln(z_1^2)$, $\ln(z_2^2)$ and $\ln((z_1-z_2)^2)$ through the ratio at short distance in coordinate space. For readers' convenience, a Mathematica notebook on the hybrid scheme matching kernel is attached to the source package on arXiv.

\subsection{Discussions and Prospect}

Before closing this section we give a few remarks in order. 
\begin{itemize}
\item  In this work, we have focused on one of the Lorentz structures of DA of a light $\Lambda$ baryon at the leading twist, but the renormalization strategy is also applicable to the other Lorentz structures at the leading twist at one loop accuracy. 

\item This renormalization strategy is likely to be generalized to  twist-4 LCDAs. Since a baryon is interpolated by three quark fields,  there are nine LCDAs at  twist-4~\cite{Braun:2000kw}.  For an explicit distribution amplitude, one needs to pick up the appropriate gamma matrices and insert them into the interpolating operator to project out the desired component. 

Furthermore, several interesting works~\cite{Bhattacharya:2020xlt,Bhattacharya:2020jfj} have recently calculated the perturbative contributions to PDFs at  twist-3 in LaMET, and pointed out that after properly taking into account the zero-mode contributions (or the so-called endpoint contributions), the  twist-3  equal-time matrix elements can be used to extract the corresponding lightcone PDFs. Accordingly, future applications to twist-4 LCDAs of a light baryon should also investigate the zero-mode contributions. 

\item The ratio scheme is used to remove the UV divergences in small separations which is not always optimal. In particular, if there is mixing between matrix elements, an RI-MOM scheme (e.g., \cite{Constantinou:2013pba, Constantinou:2017sej}) is likely beneficial. When there is mixing between matrix elements, the RI/MOM scheme can use the quark matrix elements and project out the contributions. This has been used in the analysis of parton distribution functions~\cite{Green:2017xeu,Chen:2017mie,Chen:2017mzz,Alexandrou:2017huk,Chen:2018xof,Liu:2018tox,Izubuchi:2018srq,Zhang:2018nsy,Alexandrou:2018pbm,Alexandrou:2019lfo}.

However, the implementation of RI/MOM scheme in the baryon DA case is not straightforward. The interpolating current of a baryon is made of three quarks, which does not have any corresponding quark transition matrix elements that can be constructed on the lattice. In addition, the baryon DA is a two-dimensional distribution, and the use of discrete symmetries to constrain the mixing pattern is also more complicated than the one-dimensional PDF case.

\end{itemize}

\section{Summary}

To summarize, this paper continues the work on a direct method to extract the shape distribution of LCDA of a light baryon through the simulation of equal-time correlation functions under the framework of large momentum effective theory. To clear the obstacles on the road to renormalize the quasi-DAs, we have developed a hybrid renormalization scheme. 
By combining the self-renormalization at large spatial separations and the ratio scheme at short spatial separations, the hybrid renormalization scheme removes the UV divergences without introducing extra nonperturbative effects.  The corresponding equal-time correlation functions have been calculated in the coordinate space to the next-to-leading order, and the matching kernel between the quasi-DAs and LCDAs has been derived.

In the calculation, the $\Lambda $ baryon has been taken as an example to demonstrate the scheme, but such a scheme can be straightforwardly generalized to other octet and decuplet baryons. 
This approach offers a practical methodology for computing LCDA which can be carried out in lattice calculations, and a preliminary analysis using this approach is now being conducted by members of Larton Parton collaboration. 

\begin{acknowledgments}
We would like to thank  Jun Zeng, Zhifu Deng, Minhuan Chu, Jun Hua, and Qi-An Zhang for their insightful comments and invaluable discussions.
This work is supported in part by the Natural Science Foundation of China under Grants No.12125503, No. U2032102, No. 12061131006,  and No. 12335003. Part of the computations in this paper was run on the Siyuan-1 cluster supported by the Center for High Performance Computing at Shanghai Jiao Tong University.
\end{acknowledgments}

\appendix
\section{One-loop calculation of the spatial correlator in coordinate space }\label{sec:LBDA_one_loop}

In this section, we will present one-loop results for the spatial correlator in dimensional regularization with $\overline{\text{MS}}$ renormalization. 
The results are gauge invariant, and the Feynman gauge will be adopted in the practical calculation.

As shown in Fig.~\ref{Pspic}, there are twelve distinct diagrams to calculate which can be divided into three categories: quark-quark, quark-Wilson line, and Wilson line-Wilson line.

We take Fig.~\ref{Pspic}(e)  as the example to illustrate the calculation, in which the one-loop corrections are
 \begin{equation} 
 \widetilde{O}_e=
 \left(\psi _1 \left(z_1\right) \left(i g_s \int d^d \eta _1 \bar \psi _1 \left(\eta _1\right) \slashed A \left(\eta _1\right) \psi _1 \left(\eta _1\right)\right)\right){}^T 
 \left(-i g_s \int_0^1 d t_1 z_1 \cdot A \left(t_1 z_1\right)\right) \Gamma  \psi _2 \left(z_2\right) \psi _3 (0) . 
 \end{equation} 
The color indexes and the parameter $\left(\displaystyle\frac{\mu ^2}{e^{\ln (4 \pi )-\gamma_E}}\right)^{\epsilon }$ are not written out explicitly.
The gluon and quark propagators in  the coordinate space are 
\begin{eqnarray} 
 G(x-y)=\frac{\Gamma(d / 2-1)}{4 \pi^{d / 2}} \frac{-g_{\mu \nu}}{\left(-(x-y)^2+i \epsilon\right)^{d / 2-1}},\\
Q(x-y)=\frac{\Gamma(d / 2)}{2 \pi^{d / 2}} \frac{i(\slashed x-\slashed y)}{\left(-(x-y)^2+i \epsilon\right)^{d / 2}}. 
\end{eqnarray}

Following the tedious but standard routine: Schwinger parameterization, plus field Fourier transformation, completing the square, parameter shifting, Dirac algebra simplification, and parameter integration, one arrives at
\begin{align}
 &\widetilde{O}_e= g_s^2 \frac{(-i)^{d / 2-1}}{8 \pi^{d / 2}} \int d^d k_1 \int_0^1 d t_1 \int_0^\infty d \sigma_1 \int_0^\infty  d\sigma_2 \sigma_1^{d / 2-1} \sigma_2^{d / 2-2}\left(\sigma_1+\sigma_2\right)^{-d / 2} 
 \\
 &\times e^{\frac{i\left(4 \left(\sigma_1+\sigma_2 t_1\right)z_1(k_1 \cdot n_z)+k_1^2-4 \sigma_1 \sigma_2\left(t_1-1\right)^2 (-z_1^2))\right)}{4\left(\sigma_1+\sigma_2\right)}}  \psi_1^T\left(k_1\right) \left((-z_1^2)-\frac{z_1(k_1 \cdot n_z)+\left(\sigma_1+\sigma_2 t_1\right)(-z_1^2)}{\sigma_1+\sigma_2}\right) \Gamma\psi_2\left(z_2\right) \psi_3(0),\notag
\end{align}    
where $\sigma_1$ and $\sigma_2$ are Schwinger parameters, and $k_1$ is from the Fourier transformation of $\psi_1^T(z_1)$.
Note that terms like  $k_1^2$ or $\slashed k_1 \psi (z)$ have been neglected in the calculation due to the equation of motion. 
By changing $(\sigma_1,\sigma_2)$ to $(\sigma,\eta)$ with $\sigma_1=\displaystyle\frac{\sigma}{\eta}$ and $\sigma_2=\displaystyle\frac{\sigma}{1-\eta}$, the above result can be rearranged as
\begin{eqnarray}
\widetilde{O}_e  &=&-g_s^2 \frac{(-i)^{d / 2-1}}{8 \pi^{d / 2}} \int d^d k_1 \int_0^1 d t_1 \int_0^1 d \eta \int_0^\infty d \sigma \sigma^{\frac{d}{2}-3}
\left(\left(1-\eta\right)z_1 (k_1 \cdot n_z ) +\sigma\left(t_1-1\right) (-z_1^2)\right)
\nonumber \\
&&\times e^{-i z_1 n_z \cdot \left(k_1\left(-\left(\eta\left(t_1-1\right)\right)-1\right)+\sigma\left(t_1-1\right)^2 z_1 n_z \right)}  \psi_1^T\left(k_1\right)\Gamma\psi_2\left(z_2\right) \psi_3(0), \label{OeA}
\end{eqnarray} 
and then one can calculate the two parts.

For the term involving $(-z_1^2)$ in Eq.~\eqref{OeA}, we further define
\begin{equation}
\begin{aligned}
 \widetilde{O}_{e1}=&-g_s^2 \frac{1}{8 \pi^{d / 2}} \Gamma(d / 2-1) \int d^d k_1 \int_0^1 d t_1 \int_0^1 d \eta \left(1-t_1\right)^{3-d}\left(z_1^2\right)^{2-d / 2} 
 \\&
 \times e^{i z_1 k_1\left(\eta\left(t_1-1\right)+1\right)} \psi_1^T\left(k_1\right) \Gamma  \psi_2\left(z_2\right) \psi_3(0),
\end{aligned}
\end{equation}
and we can have the simplified form
\begin{equation}
\begin{aligned}
 \widetilde{O}_{e1}=-g_s^2 \frac{1}{8 \pi^{d / 2}} \Gamma(d / 2-1)\left(z_1^2\right)^{2-d / 2} &\int_0^1 d t_0 \int_0^1 d \eta\left(t_0\right)^{3-d} 
   \psi_1^T\left(\left(1-\eta t_0\right) z_1\right) \Gamma \psi_2\left(z_2\right) \psi_3(0),
\end{aligned}
\end{equation}   
with $t_0=1-t_1$. 
The $t_0\to 0$ corresponds to a UV divergence since that divergence is regularized by $d<4$ and one end of the Wilson line approaches $z_1$ when $t_0\to 0$.
One can separate this divergence from the rest  by using $\psi^T((1-\eta t_0)z_1)=\left(\psi^T((1-\eta t_0) z_1)-\psi^T(z_1)\right )+\psi^T(z_1)$.
Then it is straightforward to obtain the results for  these two parts  
 \begin{eqnarray} 
 \widetilde{O}_{e11}=& \displaystyle\frac{\alpha _s C_F}{4 \pi } \left(\frac{1}{\epsilon _{\text{UV}}}+\log \left(\frac{1}{4} \mu ^2 z_1^2 e^{2 \gamma_E }\right)\right) 
 \psi _1^T \left(z_1\right) \Gamma  \psi _2 \left(z_2\right) \psi _3 (0), \\
 \widetilde{O}_{e12}=&\displaystyle \frac{\alpha _s C_F}{2 \pi } \int_0^1 d \eta  \left(\frac{1-\eta }{\eta }\right)  _+ 
 (\psi _1^T \left((1-\eta ) z_1\right) \Gamma  \psi _2 \left(z_2\right) \psi _3 (0). 
 \end{eqnarray}

For the $z_1(k_1 \cdot n_z)$ term in Eq.~\eqref{OeA}
\begin{equation}
\begin{aligned}
 \widetilde{O}_{e2}=&-g_s^2 \frac{(-i)^{d / 2-1}}{8 \pi^{d / 2}} \int d^d k_1 \int_0^1 d t_1 \int_0^1 d \eta \int_0^\infty d\sigma \sigma^{\frac{d}{2}-3}\left(1-\eta\right) z_1(k_1 \cdot n_z) 
  \\&
\times e^{-i z_1 n_z \cdot \left(k_1\left(-\left(\eta\left(t_1-1\right)\right)-1\right)+\sigma\left(t_1-1\right)^2 z_1 n_z\right)} 
\psi_1^T\left(k_1\right)\Gamma \psi_2\left(z_2\right) \psi_3(0),
\end{aligned}
\end{equation}  
there is an IR divergence: 
 \begin{eqnarray}
 \begin{aligned}
     \widetilde{O}_{e2}=&-C_F \frac{g_s^2}{8 \pi^2} \int_0^1 d \eta\left(\left(\ln \left(\frac{1}{4} \mu^2 z_1^2 e^{2 \gamma_E}\right)+\frac{1}{\epsilon_{\mathrm{IR}}}+2\right)\left(\frac{1-\eta}{\eta}\right)_{+}+\left(\frac{2 \ln \eta} {\eta}\right)_{+}\right) 
\\
&\times \psi_1^T\left(z_1(1-\eta)\right) \Gamma  \psi_2\left(z_2\right) \psi_3(0).
 \end{aligned}
\end{eqnarray}

Collecting all these pieces and removing the UV divergence in the $\overline{\rm MS}$ scheme give the final result: 
 \begin{equation} 
 \begin{aligned} 
 \widetilde{O}_e&=
 \frac{\alpha_s C_F}{4 \pi} \ln \left(\frac{1}{4}\mu_{\text{UV}} ^2 z_1^2 e^{2 \gamma_E}\right)  \psi _1^T \left(z_1\right) \Gamma  \psi _2 \left(z_2\right) \psi _3 (0) 
 \\& 
-\frac{\alpha_s C_F }{2 \pi} \int_0^1 d \eta \left(\frac{1-\eta }{\eta } \right) _+ \left(\ln \left(\frac{1}{4}\mu_{\text{IR}} ^2 z_1^2 e^{2 \gamma_E }\right)+\frac{1}{\epsilon _{\text{IR}}}+1\right) \psi _1^T \left((1-\eta ) z_1\right) \Gamma  \psi _2 \left(z_2\right) \psi _3 (0) 
 \\& 
-\frac{\alpha_s C_F}{ \pi} \int_0^1 d \eta \left(\frac{\ln \eta }{\eta } \right) _+ \psi _1^T \left((1-\eta ) z_1\right) \Gamma  \psi _2 \left(z_2\right) \psi _3 (0), 
 \end{aligned} 
 \end{equation}
where $\displaystyle\alpha_s=\frac{g^2_s}{4 \pi}$. We have checked that after making a Fourier transformation, the above results are consistent with Ref.~\cite{Deng:2023csv}  in momentum space. 

Then, in the same manner, results for the  quark-Wilson-line diagrams are derived as: 

\begin{align}
	 \widetilde{O}_{d}&=
   \frac{\alpha_s C_F}{8}\left(L_{12}^{\text{UV}}-L_{1}^{\text{UV}}\right)\psi_1^T\left( z_1\right) \Gamma \psi_2\left(z_2\right) \psi_3(0)
\notag\\
  &-\frac{\alpha_s C_F }{4 \pi}  \int_0^1 d \eta
   \psi_1^T\left((1-\eta) z_1+\eta z_2\right) \Gamma \psi_2\left(z_2\right) \psi_3(0)
    \left\{
    \left( L_{12}^{\text{IR}}+1+\frac{1}{\epsilon_{\mathrm{IR}}}\right) 
    \left(\frac{1-\eta}{\eta}\right)_{+}
    +2\left(\frac{\ln \eta}{\eta}\right)_+ \right\}
\notag\\
    &+\frac{\alpha_s C_F }{4 \pi}  \int_0^1 d \eta
    \psi_1^{T}\left((1-\eta) z_1\right) \Gamma \psi_2\left(z_2\right) \psi_3(0)
    \left\{
    \left( L_1^{\text{IR}}+1+\frac{1}{\epsilon_{\mathrm{IR}}}\right) 
    \left(\frac{1-\eta}{\eta}\right)_{+}
    +2\left(\frac{\ln \eta}{\eta}\right)_+ \right\}, 
\end{align} 


\begin{align}
  \widetilde{O}_{g}&=
   \frac{\alpha_s C_F}{8}\left(L_{12}^{\text{UV}}-L_{2}^{\text{UV}}\right)\psi_1^T\left( z_1\right) \Gamma \psi_2\left(z_2\right) \psi_3(0)
\notag\\
  &
  -\frac{\alpha_s C_F }{4 \pi} \int_0^1 d \eta
  \psi_1^{T}\left(z_1\right) \Gamma \psi_2\left(\eta z_1+(1-\eta) z_2\right) \psi_2\left(z_2\right) \psi_3(0) 
    \left\{
    \left( L_{12}^{\text{IR}}+1+\frac{1}{\epsilon_{\mathrm{IR}}} \right)
    \left(\frac{1-\eta}{\eta}\right)_{+}
    +2\left(\frac{\ln \eta}{\eta}\right)_+ \right\}
\notag\\
  &+\frac{\alpha_s C_F }{4 \pi} \int_0^1 d \eta
      \psi_1^{T}\left(z_1\right) \Gamma \psi_2\left((1-\eta) z_2\right) \psi_3(0)
    \left\{
    \left( L_2^{\text{IR}}+1+\frac{1}{\epsilon_{\mathrm{IR}}} \right)
    \left(\frac{1-\eta}{\eta}\right)_{+}
    +2\left(\frac{\ln \eta}{\eta}\right)_+ \right\}, 
\end{align}

\begin{equation}
    \begin{aligned}
    \widetilde{O}_{e}&= 
    \frac{\alpha_s C_F}{4 \pi} 
    L_1^{\text{UV}}
    \psi_1^{T}\left(z_1\right) \Gamma \psi_2\left(z_2\right) \psi_3(0) 
    \\
    & -\frac{\alpha_s C_F}{2 \pi} \int_0^1 d \eta \psi_1^{T}\left((1-\eta) z_1\right) \Gamma \psi_2\left(z_2\right) \psi_3(0) 
    \left\{
    \left(L_1^{\text{IR}}+1+\frac{1}{\epsilon_{\mathrm{IR}}}\right)
    \left(\frac{1-\eta}{\eta}\right)_{+}
    +2\left(\frac{\ln \eta}{\eta}\right)_{+}\right\},
\end{aligned}
    \end{equation} 
\begin{equation}
    \begin{aligned}
    \widetilde{O}_{h}&= 
     \frac{\alpha_s C_F}{8 \pi} 
    L_1^{\text{UV}}
    \psi_1^{T}\left(z_1\right) \Gamma \psi_2\left(z_2\right) \psi_3(0) 
    \\
    & -\frac{\alpha_s C_F}{4 \pi} \int_0^1 d \eta \psi_1^{T}\left(z_1\right) \Gamma \psi_2\left(z_2\right) \psi_3\left(\eta z_1\right) 
    \left\{
    \left(L_1^{\text{IR}}+1+\frac{1}{\epsilon_{\mathrm{IR}}}\right)
    \left(\frac{1-\eta}{\eta}\right)_+
    +2\left(\frac{\ln \eta}{\eta}\right)_{+}\right\},
    \end{aligned}
    \end{equation} 
\begin{equation}
    \begin{aligned}
        \widetilde{O}_{f}&= 
         \frac{\alpha_s C_F}{4 \pi} 
        L_2^{\text{UV}}
        \psi_1^{T}\left(z_1\right) \Gamma \psi_2\left(z_2\right) \psi_3(0) 
        \\
        & -\frac{\alpha_s C_F}{2 \pi} \int_0^1 d \eta \psi_1^{T}\left( z_1\right) \Gamma \psi_2\left((1-\eta) z_2\right) \psi_3(0) 
        \left\{
        \left(L_2^{\text{IR}}+1+\frac{1}{\epsilon_{\mathrm{IR}}}\right)
        \left(\frac{1-\eta}{\eta}\right)_{+}
        +2\left(\frac{\ln \eta}{\eta}\right)_{+}\right\} ,
\end{aligned}
    \end{equation} 
\begin{equation}
    \begin{aligned}
        \widetilde{O}_{i}&= 
         \frac{\alpha_s C_F}{8 \pi} 
        L_2^{\text{UV}}
        \psi_1^{T}\left(z_1\right) \Gamma \psi_2\left(z_2\right) \psi_3(0) 
        \\
        & -\frac{\alpha_s C_F}{4 \pi} \int_0^1 d \eta \psi_1^{T}\left(z_1\right) \Gamma \psi_2\left(z_2\right) \psi_3\left(\eta z_2\right) 
        \left\{
        \left(L_2^{\text{IR}}+1+\frac{1}{\epsilon_{\mathrm{IR}}}\right)
        \left(\frac{1-\eta}{\eta}\right)_+
        +2\left(\frac{\ln \eta}{\eta}\right)_{+}\right\}.
        \end{aligned}
\end{equation}    
It should be mentioned that there are both IR and UV singularities in cases (e, h, f, i), while there are no UV divergences for cases~(d, g).

The last category is the Wilson line-Wilson line vertex, corresponding to Fig-~\ref{Pspic} (j, k, l).
After similar calculations, the one-loop results can be given as
 \begin{align} 
 &\widetilde{O}_{k} = \frac{\alpha _s C_F}{2 \pi }
 \left(L_1^{\text{UV}}+2\right)
 \psi_1^{T}\left( z_1\right) \Gamma \psi_2\left(z_2\right) \psi_3(0)  , 
 \\& 
\widetilde{O}_{l} = \frac{\alpha _s C_F}{2 \pi } 
\left(L_2^{\text{UV}}+2\right)
\psi_1^{T}\left( z_1\right) \Gamma \psi_2\left(z_2\right) \psi_3(0)  , 
 \\& 
\widetilde{O}_{j} = -\frac{\alpha _s C_F}{4 \pi } 
\left( 
L_1^{\text{UV}}+L_2^{\text{UV}}-L_{12}^{\text{UV}}
+2\right)
\psi_1^{T}\left( z_1\right) \Gamma \psi_2\left(z_2\right) \psi_3(0)  . 
 \end{align} 
    
Note that in these cases, only UV divergences   arise when the two ends of the gluons coincide with each other.
The cases (e, f) and cases (d, g, h, i) have different color coefficients \cite{Deng:2023csv}.
These color differences will also appear in the following cases.
More precisely, for cases (e, f, k, l), the color algebra gives the same results as the meson case, which is $C_F$.
For other cases, the color parameter is $-\displaystyle\frac{C_F}{2}$.

The last pattern, quark-quark, is shown in Fig-1 (a, b, c).
Following the same routine, these results can be written down directly
\begin{align}
	 \widetilde{O}_{a}=
  &-\frac{\alpha_s C_F}{4 \pi} 
   \int_0^1 d \eta_1 \int_0^{1-\eta_1} d \eta_2 
   \left( L_{12}^{\text{IR}}-3+\frac{1}{\epsilon_{\mathrm{IR}}}\right)
  \psi_1^{T}\left(z_1\left(1-\eta_1\right)+z_2 \eta_1\right) \Gamma \psi_2\left(z_2\left(1-\eta_2\right)+z_1 \eta_2\right) \psi_3(0),
    \\
    \widetilde{O}_{b}= & -\frac{\alpha_s C_F}{8 \pi}  \int_0^1 d \eta_1 \int_0^{1-\eta_1} d \eta_2 
    \left(L_1^{\text{IR}}-1+\frac{1}{\epsilon_{\mathrm{IR}}}\right)
    \psi_1^{T}\left((1-\eta_1) z_1\right) \Gamma \psi_2\left(z_2\right) \psi_3\left(\eta_2 z_1\right) ,
    \\
    \widetilde{O}_{c}= & -\frac{\alpha_s C_F}{8 \pi} \int_0^1 d \eta_1 \int_0^{1-\eta_1} d \eta_2 
    \left(L_2^{\text{IR}}-1+\frac{1}{\epsilon_{\mathrm{IR}}}\right)
    \psi_1^{T}\left( z_1\right) \Gamma \psi_2\left((1-\eta_1)z_2\right) \psi_3\left(\eta_2 z_2\right).
\end{align}
  
The UV divergences have been subtracted in all these cases.
It should be mentioned that case (a) has an extra finite part compared to case (b) and case (c).

Putting all these contributions together and sandwiching them between the vacuum state $\left \langle0|\right.$ and the lowest-order Fock state $\left.|uds\right \rangle$, one can obtain the results in Eq.~(\ref{eq:Mpert}).

Additionally, the one-loop correction for the local matrix elements can also be given.
In this case, only the quark-quark category needs to be considered.
Since the $z^2 $ is equal to $0$ in the local case, the local matrix elements of quasi-DA and LCDA can be obtained by a similar calculation procedure.
The corresponding local matrix element can be given as
\begin{equation}\label{eq:2local}
\begin{aligned}
& {M}_{p}(0,0,0,P^z,\mu) = \left(1 - \frac{\alpha_s C_F}{4 \pi} \frac{1}{\epsilon_{\rm IR}}\right) {M}_{0} \left(0 , 0 , 0 ,P^z,\mu\right),
\end{aligned}
\end{equation}
in which the UV divergences   have been subtracted.

\section{Explicit expressions for the hybrid counterterm}\label{sec:Integrals}

In the hybrid renormalization scheme, the counterterm in the matching kernel can be split into different regions,  such as $I_{\rm H}$, $I_{\rm HSI}$ and $I_{\rm S}$ in Eq.~(\ref{eq-f}). In this appendix, we provide the explicit expressions for these terms. Those expressions can also be found in a Mathematica notebook in the source package on arXiv.

We first give the necessary master formulas: 
\begin{equation}
\begin{aligned}
\text{I1}\left(\left\{L_2,L_1\right\},p\right) \equiv &\int_{L_1}^{L_2} \frac{d z}{2\pi} e^{i p z} \ln[z^2] 
\\
=&-\frac{i \left(2 \left(\gamma_E +\log \left(-i L_2 p\right)\right)+\left(-1+e^{i L_2 p}\right) \log \left(L_2^2\right)+2 \Gamma \left(0,-i p L_2\right)\right)}{2 \pi  p}\\
&+\frac{i \left(2 \left(\gamma_E +\log \left(-i L_1 p\right)\right)+\left(-1+e^{i L_1 p}\right) \log \left(L_1^2\right)+2 \Gamma \left(0,-i p L_1\right)\right)}{2\pi  p},
\end{aligned}
\end{equation}
\begin{equation}
\text{I0}\left(\left\{L_2,L_1\right\},p\right) 
\equiv \int_{L_1}^{L_2} \frac{d z}{2\pi} e^{i p z}
=-\frac{i \left(-1+e^{i L_2 p}\right)}{2 \pi  p}+\frac{i \left(-1+e^{i L_1 p}\right)}{2 \pi  p}.
\end{equation}
One can obtain the following results based on the above master integrals for convenience ($L>0$)
\begin{equation}
\begin{aligned}
\text{I1t}\left(\left\{L,-L\right\},p\right) 
&\equiv i[\text{I1}\left(\left\{L,0\right\},p\right) -\text{I1}\left(\left\{0,-L\right\},p\right)] 
\\&
= \frac{2 (-\text{Ci}(L p)+\log (L) \cos (L p)+\log (p)+\gamma_E )}{\pi  p},
\end{aligned}
\end{equation}
\begin{equation}
\text{I0}\left(\left\{\infty,L\right\},p\right) 
=\frac{\delta (p)}{2}+\frac{i}{2 \pi  p}+\frac{i \left(-1+e^{i L p}\right)}{2 \pi  p},
\end{equation}
\begin{equation}
\text{I0}\left(\left\{-L,-\infty\right\},p\right) 
=-\frac{i \left(-1+e^{-i L p}\right)}{2 \pi  p}+\frac{\delta (p)}{2}-\frac{i}{2 \pi  p},
\end{equation}
\begin{equation}
\text{I0}\left(\left\{\infty,-\infty\right\},p\right) 
=\delta (p).
\end{equation}
Based upon the above formulas, all the integrals in the Eq.~(\ref{eq-f}) can be written down ($z_s>0$).
\begin{align}
&I_{\rm H} [p_1,p_2] \equiv \int \frac{d z_1}{2\pi} \frac{d z_2}{2\pi} e^{i p_1 z_1 +  i p_2 z_2}  \left[ \frac{7}{8} \ln\left(z_1^2\right) + \frac{7}{8} \ln\left(z_2^2\right) + \frac{3}{4} \ln\left((z_1-z_2)^2\right)\right] \\
&\quad \quad \quad \quad \times \left(\theta(2z_s-|z_1|)\theta(z_s-|z_2|)+\theta(z_s-|z_1|)\theta(|z_2|-z_s)\theta(2z_s-|z_2|)\right) \notag\\
&= \frac{7}{8} \left[\text{I0}\left(\left\{2 z_s,-2 z_s\right\},p_2\right) \text{I1}\left(\left\{z_s,-z_s\right\},p_1\right)+\left(\text{I0}\left(\left\{2 z_s,-2
   z_s\right\},p_1\right)-\text{I0}\left(\left\{z_s,-z_s\right\},p_1\right)\right)
   \text{I1}\left(\left\{z_s,-z_s\right\},p_2\right) \right.\notag\\
   &\left.+\text{I0}\left(\left\{z_s,-z_s\right\},p_2\right) \left(\text{I1}\left(\left\{2 z_s,-2
   z_s\right\},p_1\right)-\text{I1}\left(\left\{z_s,-z_s\right\},p_1\right)\right)+\text{I0}\left(\left\{z_s,-z_s\right\},p_1\right) \text{I1}\left(\left\{2
   z_s,-2 z_s\right\},p_2\right)\right] \notag\\
&+\frac{3}{8 \pi  \left(p_1+p_2\right)}\left[ \sin \left(\left(p_1+p_2\right) z_s\right)
   \left[\text{I1}\left(\left\{z_s,-z_s\right\},p_1\right)+\text{I1}\left(\left\{z_s,-z_s\right\},-p_2\right)\right.\right.\notag\\
   &\left.\left.-\text{I1}\left(\left\{2 z_s,-2
   z_s\right\},p_1\right)-\text{I1}\left(\left\{2 z_s,-2 z_s\right\},-p_2\right)+\text{I1}\left(\left\{3 z_s,-3 z_s\right\},p_1\right)+\text{I1}\left(\left\{3
   z_s,-3 z_s\right\},-p_2\right)\right] \right.\notag\\
   &\left.+\sin \left(2 \left(p_1+p_2\right) z_s\right)
   \left[-\text{I1}\left(\left\{z_s,-z_s\right\},p_1\right)-\text{I1}\left(\left\{z_s,-z_s\right\},-p_2\right) \right.\right.\notag\\
   &\left.\left.+\text{I1}\left(\left\{3 z_s,-3
   z_s\right\},p_1\right)+\text{I1}\left(\left\{3 z_s,-3 z_s\right\},-p_2\right)\right] \right.\notag\\
   &\left.+\cos \left(2 \left(p_1+p_2\right) z_s\right)
   \left[-\text{I1t}\left(\left\{z_s,-z_s\right\},p_1\right)+\text{I1t}\left(\left\{z_s,-z_s\right\},-p_2\right) \right.\right.\notag\\
   &\left.\left.+\text{I1t}\left(\left\{3 z_s,-3
   z_s\right\},p_1\right)-\text{I1t}\left(\left\{3 z_s,-3 z_s\right\},-p_2\right)\right] \right.\notag\\
   &\left.+\cos \left(\left(p_1+p_2\right) z_s\right)
   \left[-\text{I1t}\left(\left\{z_s,-z_s\right\},p_1\right)+\text{I1t}\left(\left\{z_s,-z_s\right\},-p_2\right)-\text{I1t}\left(\left\{2 z_s,-2
   z_s\right\},p_1\right)\right.\right.\notag\\
   &\left.\left.+\text{I1t}\left(\left\{2 z_s,-2 z_s\right\},-p_2\right)+\text{I1t}\left(\left\{3 z_s,-3
   z_s\right\},p_1\right)-\text{I1t}\left(\left\{3 z_s,-3 z_s\right\},-p_2\right)\right] \right],\notag
\end{align}

\begin{align}
&I_{\rm HSI} [p_1,p_2] \equiv \int \frac{d z_1}{2\pi} \frac{d z_2}{2\pi} e^{i p_1 z_1 +  i p_2 z_2} \theta(z_s-|z_1|)\theta(|z_2|-2z_s)
\\&
\quad \quad \quad \quad \, \,\,\times\left[ \frac{7}{8} \ln\left(z_1^2\right) + \frac{7}{8} \ln\left((2 z_s)^2\right) + \frac{3}{4} \ln\left((z_1-2 z_s {\rm sign}[z_2])^2\right)\right] \notag\\
&= \frac{1}{8} \left[6 e^{2 i p_1 z_s} \text{I1}\left(\left\{-z_s,-3 z_s\right\},p_1\right) \text{I0}\left(\left\{\infty ,2 z_s\right\},p_2\right)+6 e^{-2 i p_1
   z_s} \text{I1}\left(\left\{3 z_s,z_s\right\},p_1\right) \text{I0}\left(\left\{-2 z_s,-\infty \right\},p_2\right)\right.\notag\\
   &\left.+7\left(\text{I0}\left(\{\infty ,-\infty
   \},p_2\right)-\text{I0}\left(\left\{2 z_s,-2 z_s\right\},p_2\right)\right) \left(\log \left(4 z_s^2\right)
   \text{I0}\left(\left\{z_s,-z_s\right\},p_1\right)+\text{I1}\left(\left\{z_s,-z_s\right\},p_1\right)\right)\right],\notag
\end{align}

\begin{align}
&I_{\rm HSII} [p_1,p_2] \equiv \int \frac{d z_1}{2\pi} \frac{d z_2}{2\pi} e^{i p_1 z_1 +  i p_2 z_2} \theta(|z_1|-2z_s)\theta(z_s-|z_2|)
\\&
\quad \quad \quad \quad \quad \times  \left[ \frac{7}{8} \ln\left((2 z_s)^2\right) + \frac{7}{8} \ln\left(z_2^2\right) + \frac{3}{4} \ln\left(({\rm sign}[z_1]2 z_s- z_2)^2\right)\right] \notag\\
&= \frac{1}{8} \left[6 e^{2 i p_2 z_s} \text{I1}\left(\left\{-z_s,-3 z_s\right\},p_2\right) \text{I0}\left(\left\{\infty ,2 z_s\right\},p_1\right)+6 e^{-2 i p_2
   z_s} \text{I1}\left(\left\{3 z_s,z_s\right\},p_2\right) \text{I0}\left(\left\{-2 z_s,-\infty \right\},p_1\right) \right.\notag\\
   &\left.+7\left(\text{I0}\left(\{\infty ,-\infty
   \},p_1\right)-\text{I0}\left(\left\{2 z_s,-2 z_s\right\},p_1\right)\right) \left(\log \left(4 z_s^2\right)
   \text{I0}\left(\left\{z_s,-z_s\right\},p_2\right)+\text{I1}\left(\left\{z_s,-z_s\right\},p_2\right)\right)\right],\notag
\end{align}

\begin{align}
&I_{\rm HSIII} [p_1,p_2] \equiv \int \frac{d z_1}{2\pi} \frac{d z_2}{2\pi} e^{i p_1 z_1 +  i p_2 z_2} \theta(|z_1|-z_s)\theta(|z_2|-z_s)\theta(z_s-|z_1-z_2|) 
\\&
\times \left[ \frac{7}{8} \ln \left(\left(z_s+\left(z_1-z_2\right) \theta \left(z_1-z_2\right)\right){}^2\right)+\frac{7}{8} \ln \left(\left(z_s+\left(z_2-z_1\right)
   \theta \left(z_2-z_1\right)\right){}^2\right) +\frac{3}{4} \ln \left(\left(z_1-z_2\right){}^2\right) \right] 
   \notag\\
&= \frac{1}{8} \text{I0}\left(\{\infty ,-\infty \},p_1+p_2\right) \left[7\log \left(z_s^2\right) \text{I0}\left(\left\{z_s,-z_s\right\},\frac{1}{2}
   \left(p_1-p_2\right)\right)+6\,
   \text{I1}\left(\left\{z_s,-z_s\right\},\frac{1}{2} \left(p_1-p_2\right)\right) \right.
   \notag\\
   &\left. +7e^{\frac{1}{2} i \left(p_1-p_2\right) z_s} \text{I1}\left(\left\{-z_s,-2 z_s\right\},\frac{1}{2} \left(p_1-p_2\right)\right)+7e^{-\frac{1}{2} i \left(p_1-p_2\right) z_s} \text{I1}\left(\left\{2
   z_s,z_s\right\},\frac{1}{2} \left(p_1-p_2\right)\right)\right] 
   \notag\\
   &+\frac{i e^{-i \left(p_1+p_2\right) z_s}}{16 \pi  \left(p_1+p_2\right)} \left[-7\log \left(z_s^2\right) \text{I0}\left(\left\{0,-z_s\right\},p_1\right)+7e^{2 i \left(p_1+p_2\right) z_s} \log
   \left(z_s^2\right) \text{I0}\left(\left\{0,-z_s\right\},-p_2\right) \right.
   \notag\\
   &\left.+7e^{2 i \left(p_1+p_2\right) z_s} \log \left(z_s^2\right)
   \text{I0}\left(\left\{z_s,0\right\},p_1\right)-7\log \left(z_s^2\right) \text{I0}\left(\left\{z_s,0\right\},-p_2\right)-6
   \text{I1}\left(\left\{0,-z_s\right\},p_1\right) \right.
   \notag\\
   &\left.+6 e^{2 i \left(p_1+p_2\right) z_s} \text{I1}\left(\left\{0,-z_s\right\},-p_2\right)-7e^{i p_1 z_s}
   \text{I1}\left(\left\{-z_s,-2 z_s\right\},p_1\right)+7e^{i \left(2 p_1+p_2\right) z_s} \text{I1}\left(\left\{-z_s,-2 z_s\right\},-p_2\right) \right.
   \notag\\
   &\left.+6 e^{2 i\left(p_1+p_2\right) z_s} \text{I1}\left(\left\{z_s,0\right\},p_1\right)-6 \text{I1}\left(\left\{z_s,0\right\},-p_2\right)+7e^{i \left(p_1+2 p_2\right) z_s}
   \text{I1}\left(\left\{2 z_s,z_s\right\},p_1\right)-7e^{i p_2 z_s} \text{I1}\left(\left\{2 z_s,z_s\right\},-p_2\right)\right],\notag
\end{align}

\begin{align}
&I_{\rm HSIV} [p_1,p_2] \equiv \int \frac{d z_1}{2\pi} \frac{d z_2}{2\pi} e^{i p_1 z_1 +  i p_2 z_2} \theta(|z_1|-z_s)\theta(|z_2|-z_s)\theta(z_s-|z_1+z_2|) 
\\&
\times \left[ \frac{7}{8} \ln \left(\left(z_s+\left(z_1+z_2\right) \theta \left(z_1+z_2\right)\right){}^2\right)+\frac{7}{8} \ln \left(\left(-z_s+\left(z_2+z_1\right)
   \theta \left(-z_2-z_1\right)\right){}^2\right)+\frac{3}{4} \ln \left(\left(2 z_s + |z_1+z_2|\right){}^2\right) \right]  
   \notag\\
   &=\frac{1}{16} \text{I0}\left(\{\infty ,-\infty \},\frac{1}{2} \left(p_1-p_2\right)\right) \left[7\log \left(z_s^2\right)
   \text{I0}\left(\left\{z_s,-z_s\right\},\frac{1}{2} \left(p_1+p_2\right)\right)\right.
   \notag\\
   &\left.+6 e^{i \left(p_1+p_2\right) z_s} \text{I1}\left(\left\{-2 z_s,-3z_s\right\},\frac{1}{2} \left(p_1+p_2\right)\right) +7e^{\frac{1}{2} i \left(p_1+p_2\right) z_s} \text{I1}\left(\left\{-z_s,-2 z_s\right\},\frac{1}{2}
   \left(p_1+p_2\right)\right)\right.
   \notag\\
   &\left.+7e^{-\frac{1}{2} i \left(p_1+p_2\right) z_s} \text{I1}\left(\left\{2 z_s,z_s\right\},\frac{1}{2} \left(p_1+p_2\right)\right)+6
   e^{-i \left(p_1+p_2\right) z_s} \text{I1}\left(\left\{3 z_s,2 z_s\right\},\frac{1}{2} \left(p_1+p_2\right)\right)\right] 
   \notag\\
   &+\frac{i e^{-i \left(p_1+p_2\right) z_s}}{16 \pi 
   \left(p_1-p_2\right)} \left[-7e^{2 i p_2 z_s} \log \left(z_s^2\right) \text{I0}\left(\left\{0,-z_s\right\},p_1\right)+7e^{2 i p_1 z_s} \log
   \left(z_s^2\right) \text{I0}\left(\left\{0,-z_s\right\},p_2\right)\right.
   \notag\\
   &\left.+7e^{2 i p_1 z_s} \log \left(z_s^2\right)
   \text{I0}\left(\left\{z_s,0\right\},p_1\right)-7e^{2 i p_2 z_s} \log \left(z_s^2\right) \text{I0}\left(\left\{z_s,0\right\},p_2\right)-6 e^{2 i\left(p_1+p_2\right) z_s} \text{I1}\left(\left\{-2 z_s,-3 z_s\right\},p_1\right)\right.
   \notag\\
   &\left.+6 e^{2 i \left(p_1+p_2\right) z_s} \text{I1}\left(\left\{-2 z_s,-3z_s\right\},p_2\right)-7e^{i \left(p_1+2 p_2\right) z_s} \text{I1}\left(\left\{-z_s,-2 z_s\right\},p_1\right)+7e^{i \left(2 p_1+p_2\right) z_s}
   \text{I1}\left(\left\{-z_s,-2 z_s\right\},p_2\right)\right.
   \notag\\
   &\left.+7e^{i p_1 z_s} \text{I1}\left(\left\{2 z_s,z_s\right\},p_1\right)-7e^{i p_2 z_s} \text{I1}\left(\left\{2
   z_s,z_s\right\},p_2\right)+6 \text{I1}\left(\left\{3 z_s,2 z_s\right\},p_1\right)-6 \text{I1}\left(\left\{3 z_s,2 z_s\right\},p_2\right)\right],\notag
\end{align}

\begin{align}
&I_{\rm S} [p_1,p_2] \equiv \int \frac{d z_1}{2\pi} \frac{d z_2}{2\pi} e^{i p_1 z_1 +  i p_2 z_2}  \theta(|z_1|-z_s)\theta(|z_2|-z_s)\theta(|z_1-z_2|-z_s)\theta(|z_1+z_2|-z_s) 
\\
&
\quad \quad \quad \quad \times\left[ \frac{7}{8} \ln\left(z_s^2\right) + \frac{7}{8} \ln\left((2z_s)^2\right) + \frac{3}{4} \ln\left(({\rm sign}[z_1]z_s-{\rm sign}[z_2]2z_s)^2\right)\right]
\notag\\
&=\delta \left(p_1-p_2\right) \delta \left(p_1+p_2\right) \left(\frac{7}{4} \log \left(4 z_s^4\right)+\frac{3}{4} \log \left(9 z_s^4\right)\right) 
\notag\\
& -\frac{\delta \left(p_1+p_2\right) \left(6 \log \left(z_s^2\right)+7 \log \left(4 z_s^4\right)\right) \sin \left(\frac{1}{2} \left(p_1-p_2\right) z_s\right)}{4
   \pi  \left(p_1-p_2\right)}
   \notag\\
&-\frac{\delta \left(p_1-p_2\right) \left(6 \log \left(9 z_s^2\right)+7 \log \left(4 z_s^4\right)\right) \sin \left(\frac{1}{2}
   \left(p_1+p_2\right) z_s\right)}{4 \pi  \left(p_1+p_2\right)} 
   \notag\\
& -\frac{\delta \left(p_2\right) \left(20 \log \left(z_s\right)+\log (3456)\right) \sin \left(p_1 z_s\right)}{4 \pi  p_1}-\frac{\delta \left(p_1\right) \left(20
   \log \left(z_s\right)+\log (3456)\right) \sin \left(p_2 z_s\right)}{4 \pi  p_2}
   \notag\\
&-\frac{\left(20 \log \left(z_s\right)+\log (128)\right) \left(p_1 \cos \left(\left(p_1+2 p_2\right) z_s\right)+p_2 \cos \left(\left(2 p_1+p_2\right)
   z_s\right)\right)}{8 \pi ^2 p_1 p_2 \left(p_1+p_2\right)}
   \notag\\
&+\frac{\left(20 \log \left(z_s\right)+\log (93312)\right) \left(p_1 \cos \left(\left(p_1-2 p_2\right) z_s\right)-p_2 \cos \left(2 p_1 z_s-p_2 z_s\right)\right)}{8 \pi ^2 p_1 \left(p_1-p_2\right) p_2}.\notag
\end{align}



\section{Double plus function}\label{sec:douplusfunc}
In this appendix, we demonstrate that the matching kernel at the one-loop level can be properly expressed in the form of the double plus function in Eq.~(\ref{eq:dpfhybrid}). This is equivalent to proving the following lemma: \\
\textbf{Lemma.} \textit{Consider a generalized function $G(x_1,x_2,x_{10},x_{20})$ satisfying the following properties 
\begin{equation}\label{eq:LTcon}
\begin{aligned} 
&\lim_{|x_1-x_{10}| \rightarrow +\infty}|x_1-x_{10}| G(x_1,x_2,x_{10},x_{20}) =_{a.e.} 0 ,\\
&\lim_{|x_2-x_{20}| \rightarrow +\infty}|x_2-x_{20}| G(x_1,x_2,x_{10},x_{20}) =_{a.e.} 0, \\
&\lim_{|x_3-x_{30}| \rightarrow +\infty}|x_3-x_{30}| G(x_1,x_2,x_{10},x_{20}) =_{a.e.} 0, \\
\end{aligned}
\end{equation}
\begin{equation}\label{eq:SDcon}
\begin{aligned}
&\exists \, \epsilon>0, \, \left| \int_{x_{1}-\epsilon}^{x_{1}+\epsilon} d x_{10} \, |x_1-x_{10}| G(x_1,x_2,x_{10},x_{20}) \right| < +\infty, \\
&\exists \, \epsilon>0, \, \left| \int_{x_{2}-\epsilon}^{x_{2}+\epsilon} d x_{20} \, |x_2-x_{20}| G(x_1,x_2,x_{10},x_{20}) \right| < +\infty ,\\
&\exists \, \epsilon>0, \, \left| \int_{x_{3}-\epsilon}^{x_{3}+\epsilon} d x_{30} \, |x_3-x_{30}| G(x_1,x_2,x_{10},x_{20}) \right| < +\infty,
\end{aligned}
\end{equation}
and in a specific form \begin{equation}\label{eq:Gx1x2}
    \begin{aligned}    G(x_1,x_2,x_{10},x_{20})=&A(x_1,x_2,x_{10},x_{20})+\delta(x_1-x_{10})B(x_2,x_{10},x_{20})+\delta(x_2-x_{20})C(x_1,x_{10},x_{20})\\+&\delta(x_3-x_{30})D(x_1,x_2,x_{10},x_{20})+\delta(x_1-x_{10})\delta(x_2-x_{20})E(x_{10},x_{20}),
    \end{aligned}
\end{equation}
where $x_3(x_{30})$ is shorthand for $1-x_1-x_2(1-x_{10}-x_{20})$ and $A,B,C,D,E$ can be expanded into power series with logarithmic terms without Dirac delta functions. $``a.e."$ is shorthand for ``almost everywhere" which means the results hold except for $x_1=x_{10}$, $x_2=x_{20}$ or $x_3=x_{30}$ here. The lemma states that if the integral $\int d x_1 d x_2 G(x_1,x_2,x_{10},x_{20})=0$, the generalized function $G(x_1,x_2,x_{10},x_{20})$ can be written as the double plus function defined in Eq.~(\ref{dpf}).} 

\textit{Proof.} Based on the condition that $\int d x_1 d x_2 G(x_1,x_2,x_{10},x_{20})=0$, we obtain
\begin{equation}
    \begin{aligned}
        E(x_{10},x_{20})=&-\int A(x_1,x_2,x_{10},x_{20})dx_1dx_2-\int B(x_2,x_{10},x_{20})dx_2-\int C(x_1,x_{10},x_{20})dx_1
        \\&-\int \delta(x_3-x_{30}) D(x_1,x_2,x_{10},x_{20})dx_1dx_2.
    \end{aligned}
\end{equation}
Plugging $E(x_{10},x_{20})$ into Eq.~(\ref{eq:Gx1x2}) gives:
\begin{equation}
    \begin{aligned}      &G(x_1,x_2,x_{10},x_{20})=A(x_1,x_2,x_{10},x_{20})-\delta(x_1-x_{10})\delta(x_2-x_{20})\int A(y_1,y_2,x_{10},x_{20})dy_1dy_2\\
        &+\delta(x_1-x_{10})B(x_2,x_{10},x_{20})-\delta(x_1-x_{10})\delta(x_2-x_{20})\int \delta(y_1-x_{10})B(y_2,x_{10},x_{20})dy_1 dy_2\\
        &+\delta(x_2-x_{20})C(x_1,x_{10},x_{20})-\delta(x_1-x_{10})\delta(x_2-x_{20})\int \delta(y_2-x_{20})C(y_1,x_{10},x_{20})dy_1 dy_2\\
        &+\delta(x_3-x_{30})D(x_1,x_2,x_{10},x_{20})-\delta(x_1-x_{10})\delta(x_2-x_{20})\int \delta(y_3-x_{30})D(y_1,y_2,x_{10},x_{20})dy_1dy_2\\
&=\left[A(x_1,x_2,x_{10},x_{20}) +\delta(x_1-x_{10})B(x_2,x_{10},x_{20}) +\delta(x_2-x_{20})C(x_1,x_{10},x_{20}) \right.\\ 
&\left.+\delta(x_3-x_{30})D(x_1,x_2,x_{10},x_{20})\right]_\oplus,
\end{aligned}
\end{equation}
which is the double plus function defined in Eq.~(\ref{dpf}). Because of Eqs.~(\ref{eq:LTcon}) and~(\ref{eq:SDcon}), one obtains a finite result after convoluting the generalized function $G(x_1,x_2,x_{10},x_{20})$ with a Schwartz function $\Phi(x_{10},x_{20})$, 
\begin{equation}
\int d x_{10} d x_{20} G(x_1,x_2,x_{10},x_{20}) \Phi(x_{10},x_{20}) < \infty.
\end{equation}

\bibliographystyle{unsrt}

\begin{thebibliography}{99}
\bibitem{Shih:1998pb}
H.~H.~Shih, S.~C.~Lee and H.~n.~Li,
Phys. Rev. D \textbf{59}, 094014 (1999)
doi:10.1103/PhysRevD.59.094014
[arXiv:hep-ph/9810515 [hep-ph]].

\bibitem{LHCb:2015eia}
R.~Aaij \textit{et al.} [LHCb],
Nature Phys. \textbf{11}, 743-747 (2015)
doi:10.1038/nphys3415
[arXiv:1504.01568 [hep-ex]].

\bibitem{LHCb:2015tgy}
R.~Aaij \textit{et al.} [LHCb],
JHEP \textbf{06}, 115 (2015)
[erratum: JHEP \textbf{09}, 145 (2018)]
doi:10.1007/JHEP06(2015)115
[arXiv:1503.07138 [hep-ex]].

\bibitem{LHCb:2021byf}
R.~Aaij \textit{et al.} [LHCb],
Phys. Rev. D \textbf{105}, no.5, L051104 (2022)
doi:10.1103/PhysRevD.105.L051104
[arXiv:2111.10194 [hep-ex]].

\bibitem{Chernyak:1984bm}
V.~L.~Chernyak and I.~R.~Zhitnitsky,
Nucl. Phys. B \textbf{246}, 52-74 (1984)
doi:10.1016/0550-3213(84)90114-7


\bibitem{King:1986wi}
I.~D.~King and C.~T.~Sachrajda,
Nucl. Phys. B \textbf{279}, 785-803 (1987)
doi:10.1016/0550-3213(87)90019-8

\bibitem{Braun:1999te}
V.~M.~Braun, S.~E.~Derkachov, G.~P.~Korchemsky and A.~N.~Manashov,
Nucl. Phys. B \textbf{553}, 355-426 (1999)
doi:10.1016/S0550-3213(99)00265-5
[arXiv:hep-ph/9902375 [hep-ph]].


\bibitem{Anikin:2013aka}
I.~V.~Anikin, V.~M.~Braun and N.~Offen,
Phys. Rev. D \textbf{88}, 114021 (2013)
doi:10.1103/PhysRevD.88.114021
[arXiv:1310.1375 [hep-ph]].
\bibitem{Gockeler:2008xv}
M.~Gockeler, R.~Horsley, T.~Kaltenbrunner, Y.~Nakamura, D.~Pleiter, P.~E.~L.~Rakow, A.~Schafer, G.~Schierholz, H.~Stuben and N.~Warkentin, \textit{et al.}
Phys. Rev. Lett. \textbf{101}, 112002 (2008)
doi:10.1103/PhysRevLett.101.112002
[arXiv:0804.1877 [hep-lat]].

\bibitem{QCDSF:2008qtn}
V.~M.~Braun \textit{et al.} [QCDSF],
Phys. Rev. D \textbf{79}, 034504 (2009)
doi:10.1103/PhysRevD.79.034504
[arXiv:0811.2712 [hep-lat]].

\bibitem{Bali:2015ykx}
G.~S.~Bali, V.~M.~Braun, M.~G\"ockeler, M.~Gruber, F.~Hutzler, A.~Sch\"afer, R.~W.~Schiel, J.~Simeth, W.~S\"oldner and A.~Sternbeck, \textit{et al.}
JHEP \textbf{02}, 070 (2016)
doi:10.1007/JHEP02(2016)070
[arXiv:1512.02050 [hep-lat]].

\bibitem{RQCD:2019hps}
G.~S.~Bali \textit{et al.} [RQCD],
Eur. Phys. J. A \textbf{55}, no.7, 116 (2019)
doi:10.1140/epja/i2019-12803-6
[arXiv:1903.12590 [hep-lat]].

\bibitem{Deng:2023csv}
Z.~F.~Deng, C.~Han, W.~Wang, J.~Zeng and J.~L.~Zhang,
JHEP \textbf{07}, 191 (2023)
doi:10.1007/JHEP07(2023)191
[arXiv:2304.09004 [hep-ph]].

\bibitem{Ji:2013dva}
X.~Ji,
Phys. Rev. Lett. \textbf{110}, 262002 (2013)
doi:10.1103/PhysRevLett.110.262002
[arXiv:1305.1539 [hep-ph]].

\bibitem{Ji:2014gla}
X.~Ji,
Sci. China Phys. Mech. Astron. \textbf{57}, 1407-1412 (2014)
doi:10.1007/s11433-014-5492-3
[arXiv:1404.6680 [hep-ph]].

\bibitem{Cichy:2018mum}
K.~Cichy and M.~Constantinou,
Adv. High Energy Phys. \textbf{2019}, 3036904 (2019)
doi:10.1155/2019/3036904
[arXiv:1811.07248 [hep-lat]].

\bibitem{Zhao:2018fyu}
Y.~Zhao,
Int. J. Mod. Phys. A \textbf{33}, no.36, 1830033 (2019)
doi:10.1142/S0217751X18300338
[arXiv:1812.07192 [hep-ph]].

\bibitem{Ji:2020ect}
X.~Ji, Y.~S.~Liu, Y.~Liu, J.~H.~Zhang and Y.~Zhao,
Rev. Mod. Phys. \textbf{93}, no.3, 035005 (2021)
doi:10.1103/RevModPhys.93.035005
[arXiv:2004.03543 [hep-ph]].


\bibitem{Martinelli:1994ty}
G.~Martinelli, C.~Pittori, C.~T.~Sachrajda, M.~Testa and A.~Vladikas,
Nucl. Phys. B \textbf{445}, 81-108 (1995)
doi:10.1016/0550-3213(95)00126-D
[arXiv:hep-lat/9411010 [hep-lat]].

\bibitem{Zhang:2020rsx}
K.~Zhang \textit{et al.} [\ensuremath{\chi}QCD],
Phys. Rev. D \textbf{104}, no.7, 074501 (2021)
doi:10.1103/PhysRevD.104.074501
[arXiv:2012.05448 [hep-lat]].

\bibitem{LatticePartonCollaborationLPC:2021xdx}
Y.~K.~Huo \textit{et al.} [Lattice Parton Collaboration (LPC)],
Nucl. Phys. B \textbf{969}, 115443 (2021)
doi:10.1016/j.nuclphysb.2021.115443
[arXiv:2103.02965 [hep-lat]].

\bibitem{Ji:2020brr}
X.~Ji, Y.~Liu, A.~Sch\"afer, W.~Wang, Y.~B.~Yang, J.~H.~Zhang and Y.~Zhao,
Nucl. Phys. B \textbf{964}, 115311 (2021)
doi:10.1016/j.nuclphysb.2021.115311
[arXiv:2008.03886 [hep-ph]].

\bibitem{Hua:2020gnw}
J.~Hua \textit{et al.} [Lattice Parton],
Phys. Rev. Lett. \textbf{127}, no.6, 062002 (2021)
doi:10.1103/PhysRevLett.127.062002
[arXiv:2011.09788 [hep-lat]].

\bibitem{Gao:2021dbh}
X.~Gao, A.~D.~Hanlon, S.~Mukherjee, P.~Petreczky, P.~Scior, S.~Syritsyn and Y.~Zhao,
Phys. Rev. Lett. \textbf{128}, no.14, 142003 (2022)
doi:10.1103/PhysRevLett.128.142003
[arXiv:2112.02208 [hep-lat]].

\bibitem{LatticeParton:2022zqc}
J.~Hua \textit{et al.} [Lattice Parton],
Phys. Rev. Lett. \textbf{129}, no.13, 132001 (2022)
doi:10.1103/PhysRevLett.129.132001
[arXiv:2201.09173 [hep-lat]].

\bibitem{Chou:2022drv}
C.~Y.~Chou and J.~W.~Chen,
Phys. Rev. D \textbf{106}, no.1, 014507 (2022)
doi:10.1103/PhysRevD.106.014507
[arXiv:2204.08343 [hep-lat]].

\bibitem{Hua:2022wop}
J.~Hua \textit{et al.} [Lattice Parton (LPC)],
PoS \textbf{LATTICE2021}, 322 (2022)
doi:10.22323/1.396.0322

\bibitem{Gao:2022iex}
X.~Gao, A.~D.~Hanlon, N.~Karthik, S.~Mukherjee, P.~Petreczky, P.~Scior, S.~Shi, S.~Syritsyn, Y.~Zhao and K.~Zhou,
Phys. Rev. D \textbf{106}, no.11, 114510 (2022)
doi:10.1103/PhysRevD.106.114510
[arXiv:2208.02297 [hep-lat]].

\bibitem{LatticeParton:2022xsd}
F.~Yao \textit{et al.} [Lattice Parton],
[arXiv:2208.08008 [hep-lat]].

\bibitem{Su:2022fiu}
Y.~Su, J.~Holligan, X.~Ji, F.~Yao, J.~H.~Zhang and R.~Zhang,
Nucl. Phys. B \textbf{991}, 116201 (2023)
doi:10.1016/j.nuclphysb.2023.116201
[arXiv:2209.01236 [hep-ph]].

\bibitem{Ji:2022ezo}
X.~Ji,
[arXiv:2209.09332 [hep-lat]].

\bibitem{Gao:2022ytj}
X.~Gao, A.~D.~Hanlon, S.~Mukherjee, P.~Petreczky, P.~Scior, S.~Syritsyn and Y.~Zhao,
PoS \textbf{LATTICE2022}, 104 (2023)
doi:10.22323/1.430.0104

\bibitem{Gao:2022uhg}
X.~Gao, A.~D.~Hanlon, J.~Holligan, N.~Karthik, S.~Mukherjee, P.~Petreczky, S.~Syritsyn and Y.~Zhao,
Phys. Rev. D \textbf{107}, no.7, 074509 (2023)
doi:10.1103/PhysRevD.107.074509
[arXiv:2212.12569 [hep-lat]].

\bibitem{Ji:2022thb}
Y.~Ji, F.~Yao and J.~H.~Zhang,
[arXiv:2212.14415 [hep-ph]].

\bibitem{Zhang:2023tnc}
R.~Zhang,
doi:10.13016/dspace/gggm-uvss

\bibitem{Holligan:2023rex}
J.~Holligan, X.~Ji, H.~W.~Lin, Y.~Su and R.~Zhang,
Nucl. Phys. B \textbf{993}, 116282 (2023)
doi:10.1016/j.nuclphysb.2023.116282
[arXiv:2301.10372 [hep-lat]].

\bibitem{Zhang:2023bxs}
R.~Zhang, J.~Holligan, X.~Ji and Y.~Su,
Phys. Lett. B \textbf{844}, 138081 (2023)
doi:10.1016/j.physletb.2023.138081
[arXiv:2305.05212 [hep-lat]].

\bibitem{Gao:2023lny}
X.~Gao, W.~Y.~Liu and Y.~Zhao,
[arXiv:2306.14960 [hep-ph]].

\bibitem{Izubuchi:2018srq}
T.~Izubuchi, X.~Ji, L.~Jin, I.~W.~Stewart and Y.~Zhao,
Phys. Rev. D \textbf{98}, no.5, 056004 (2018)
doi:10.1103/PhysRevD.98.056004
[arXiv:1801.03917 [hep-ph]].

\bibitem{Orginos:2017kos}
K.~Orginos, A.~Radyushkin, J.~Karpie and S.~Zafeiropoulos,
Phys. Rev. D \textbf{96}, no.9, 094503 (2017)
doi:10.1103/PhysRevD.96.094503
[arXiv:1706.05373 [hep-ph]].

\bibitem{Radyushkin:2017cyf}
A.~V.~Radyushkin,
Phys. Rev. D \textbf{96}, no.3, 034025 (2017)
doi:10.1103/PhysRevD.96.034025
[arXiv:1705.01488 [hep-ph]].

\bibitem{Radyushkin:2017lvu}
A.~V.~Radyushkin,
Phys. Lett. B \textbf{781}, 433-442 (2018)
doi:10.1016/j.physletb.2018.04.023
[arXiv:1710.08813 [hep-ph]].

\bibitem{Chen:2016fxx}
J.~W.~Chen, X.~Ji and J.~H.~Zhang,
Nucl. Phys. B \textbf{915}, 1-9 (2017)
doi:10.1016/j.nuclphysb.2016.12.004
[arXiv:1609.08102 [hep-ph]].

\bibitem{Ji:2017oey}
X.~Ji, J.~H.~Zhang and Y.~Zhao,
Phys. Rev. Lett. \textbf{120}, no.11, 112001 (2018)
doi:10.1103/PhysRevLett.120.112001
[arXiv:1706.08962 [hep-ph]].

\bibitem{Ishikawa:2017faj}
T.~Ishikawa, Y.~Q.~Ma, J.~W.~Qiu and S.~Yoshida,
Phys. Rev. D \textbf{96}, no.9, 094019 (2017)
doi:10.1103/PhysRevD.96.094019
[arXiv:1707.03107 [hep-ph]].

\bibitem{Green:2017xeu}
J.~Green, K.~Jansen and F.~Steffens,
Phys. Rev. Lett. \textbf{121}, no.2, 022004 (2018)
doi:10.1103/PhysRevLett.121.022004
[arXiv:1707.07152 [hep-lat]].

\bibitem{Ji:1995tm}
X.~D.~Ji,
[arXiv:hep-ph/9507322 [hep-ph]].

\bibitem{Beneke:1998ui}
M.~Beneke,
Phys. Rept. \textbf{317}, 1-142 (1999)
doi:10.1016/S0370-1573(98)00130-6
[arXiv:hep-ph/9807443 [hep-ph]].

\bibitem{Bauer:2011ws}
C.~Bauer, G.~S.~Bali and A.~Pineda,
Phys. Rev. Lett. \textbf{108}, 242002 (2012)
doi:10.1103/PhysRevLett.108.242002
[arXiv:1111.3946 [hep-ph]].

\bibitem{Bali:2013pla}
G.~S.~Bali, C.~Bauer, A.~Pineda and C.~Torrero,
Phys. Rev. D \textbf{87}, 094517 (2013)
doi:10.1103/PhysRevD.87.094517
[arXiv:1303.3279 [hep-lat]].

\bibitem{Karbstein:2018mzo}
F.~Karbstein, M.~Wagner and M.~Weber,
Phys. Rev. D \textbf{98}, no.11, 114506 (2018)
doi:10.1103/PhysRevD.98.114506
[arXiv:1804.10909 [hep-ph]].


\bibitem{Braun:2000kw}
V.~Braun, R.~J.~Fries, N.~Mahnke and E.~Stein,
Nucl. Phys. B \textbf{589}, 381-409 (2000)
[erratum: Nucl. Phys. B \textbf{607}, 433-433 (2001)]
doi:10.1016/S0550-3213(00)00516-2
[arXiv:hep-ph/0007279 [hep-ph]].


\bibitem{Bhattacharya:2020xlt}
S.~Bhattacharya, K.~Cichy, M.~Constantinou, A.~Metz, A.~Scapellato and F.~Steffens,
Phys. Rev. D \textbf{102}, no.3, 034005 (2020)
[erratum: Phys. Rev. D \textbf{108}, no.3, 039901 (2023)]
doi:10.1103/PhysRevD.102.034005
[arXiv:2005.10939 [hep-ph]].


\bibitem{Bhattacharya:2020jfj}
S.~Bhattacharya, K.~Cichy, M.~Constantinou, A.~Metz, A.~Scapellato and F.~Steffens,
Phys. Rev. D \textbf{102}, 114025 (2020)
doi:10.1103/PhysRevD.102.114025
[arXiv:2006.12347 [hep-ph]].


\bibitem{Constantinou:2013pba}
M.~Constantinou, M.~Costa and H.~Panagopoulos,
Phys. Rev. D \textbf{88}, 034504 (2013)
doi:10.1103/PhysRevD.88.034504
[arXiv:1305.1870 [hep-lat]].

\bibitem{Constantinou:2017sej}
M.~Constantinou and H.~Panagopoulos,
Phys. Rev. D \textbf{96}, no.5, 054506 (2017)
doi:10.1103/PhysRevD.96.054506
[arXiv:1705.11193 [hep-lat]].

\bibitem{Chen:2017mzz}
J.~W.~Chen, T.~Ishikawa, L.~Jin, H.~W.~Lin, Y.~B.~Yang, J.~H.~Zhang and Y.~Zhao,
Phys. Rev. D \textbf{97}, no.1, 014505 (2018)
doi:10.1103/PhysRevD.97.014505
[arXiv:1706.01295 [hep-lat]].

\bibitem{Alexandrou:2017huk}
C.~Alexandrou, K.~Cichy, M.~Constantinou, K.~Hadjiyiannakou, K.~Jansen, H.~Panagopoulos and F.~Steffens,
Nucl. Phys. B \textbf{923}, 394-415 (2017)
doi:10.1016/j.nuclphysb.2017.08.012
[arXiv:1706.00265 [hep-lat]].



\bibitem{Chen:2017mie}
J.~W.~Chen \textit{et al.} [LP3],
Chin. Phys. C \textbf{43}, no.10, 103101 (2019)
doi:10.1088/1674-1137/43/10/103101
[arXiv:1710.01089 [hep-lat]].


\bibitem{Chen:2018xof}
J.~W.~Chen, L.~Jin, H.~W.~Lin, Y.~S.~Liu, Y.~B.~Yang, J.~H.~Zhang and Y.~Zhao,
[arXiv:1803.04393 [hep-lat]].


\bibitem{Liu:2018tox}
Y.~S.~Liu, W.~Wang, J.~Xu, Q.~A.~Zhang, S.~Zhao and Y.~Zhao,
Phys. Rev. D \textbf{99}, no.9, 094036 (2019)
doi:10.1103/PhysRevD.99.094036
[arXiv:1810.10879 [hep-ph]].


\bibitem{Zhang:2018nsy}
J.~H.~Zhang, J.~W.~Chen, L.~Jin, H.~W.~Lin, A.~Sch\"afer and Y.~Zhao,
Phys. Rev. D \textbf{100}, no.3, 034505 (2019)
doi:10.1103/PhysRevD.100.034505
[arXiv:1804.01483 [hep-lat]].


\bibitem{Alexandrou:2018pbm}
C.~Alexandrou, K.~Cichy, M.~Constantinou, K.~Jansen, A.~Scapellato and F.~Steffens,
Phys. Rev. Lett. \textbf{121}, no.11, 112001 (2018)
doi:10.1103/PhysRevLett.121.112001
[arXiv:1803.02685 [hep-lat]].


\bibitem{Alexandrou:2019lfo}
C.~Alexandrou, K.~Cichy, M.~Constantinou, K.~Hadjiyiannakou, K.~Jansen, A.~Scapellato and F.~Steffens,
Phys. Rev. D \textbf{99}, no.11, 114504 (2019)
doi:10.1103/PhysRevD.99.114504
[arXiv:1902.00587 [hep-lat]].

\bibitem{Alexandrou:2020qtt}
C.~Alexandrou, K.~Cichy, M.~Constantinou, J.~R.~Green, K.~Hadjiyiannakou, K.~Jansen, F.~Manigrasso, A.~Scapellato and F.~Steffens,
Phys. Rev. D \textbf{103}, 094512 (2021)
doi:10.1103/PhysRevD.103.094512
[arXiv:2011.00964 [hep-lat]].


\bibitem{Lin:2020fsj}
H.~W.~Lin, J.~W.~Chen and R.~Zhang,
[arXiv:2011.14971 [hep-lat]].


\end{thebibliography}

\end{document}